\documentclass[pre,aps]{revtex4}
\newcommand{\bec}[1]{\mbox{\boldmath $ #1$}}

\begin{document}
\title{THE MEAN ELECTROMOTIVE FORCE FOR MHD TURBULENCE:
THE CASE OF A WEAK MEAN MAGNETIC FIELD AND SLOW ROTATION}
\author{KARL-HEINZ R\"{A}DLER}
\affiliation{Astrophysikalisches Institut Potsdam, \\
An der Sternwarte 16, D-14482, Potsdam, Germany}
\author{NATHAN KLEEORIN}
\author{IGOR ROGACHEVSKII}
\affiliation{Department of Mechanical Engineering, The Ben-Gurion
University of the Negev, \\
POB 653, Beer-Sheva 84105, Israel}

\maketitle

The mean electromotive force that occurs in the framework of
mean--field magnetohydrodynamics is studied for cases in which
magnetic field fluctuations are not only due to the action of
velocity fluctuations on the mean magnetic field. The possibility
of magnetic field fluctuations independent of a mean magnetic
field, as they may occur as a consequence of a small--scale
dynamo, is taken into account. Particular attention is payed to
the effect of a mean rotation of the fluid on the mean
electromotive force, although only small rotation rates are
considered. Anisotropies of the turbulence due to gradients of its
intensity or its helicity are admitted. The mean magnetic field is
considered to be weak enough to exclude quenching effects. A
$\tau$-approximation is used in the equation describing the
deviation of the cross--helicity tensor from that for zero mean
magnetic field, which applies in the limit of large hydrodynamic
Reynolds numbers.

For the effects described by the mean electromotive force like
$\alpha$--effect, turbulent diffusion of magnetic fields etc. in
addition to the contributions determined by the velocity
fluctuations also those determined by the magnetic field
fluctuations independent of the mean magnetic field are derived.
Several old results are confirmed, partially under
more general assumptions, and quite a few new ones are given.
Provided the kinematic helicity and the current helicity
of the fluctuations have the same signs the $\alpha$--effect is
always diminished by the magnetic fluctuations. In the absence of
rotation these have, however, no influence on the turbulent
diffusion. Besides the diamagnetic effect due to a gradient of the
intensity of the velocity fluctuations there is a paramagnetic
effect due to a gradient of the intensity of the magnetic
fluctuations. In the absence of rotation these two effects
compensate each other in the case of equipartition of the kinetic
and magnetic energies of the fluctuations of the original turbulence,
i.e. that with zero mean magnetic field, but the rotation makes
the situation more complex. The ${\bf \Omega \times J}$--effect
works in the same way with velocity fluctuations and magnetic
field fluctuations. A contribution to the electromotive force
connected with the symmetric parts of the gradient tensor of the
mean magnetic field, which does not occur in the absence of
rotation, was found in the case of rotation, resulting from
velocity or magnetic fluctuations.

The implications of the results for the mean electromotive force
for mean--field dynamo models are discussed with special emphasis
to dynamos working without $\alpha$-effect.

The results for the coefficients defining the mean electromotive
force which are determined by the velocity fluctuations
in the case of vanishing mean motion
agree formally with the results obtained in the kinematic approach,
specified by second--order approximation and high--conductivity
limit. However, their range of validity is clearly larger.

\bigskip


\section*{1. INTRODUCTION}

The mean--field approach proved to be very useful in studying
dynamo processes in turbulently moving electrically conducting
fluids (see, e.g., Moffatt 1978, Parker 1979, Krause \& R\"{a}dler
1980, Zeldovich et al. 1983). The crucial point of this approach
is the mean electromotive force which is determined by the
fluctuations of the fluid velocity and the magnetic field. It
describes several physical effects like $ \alpha $-effect,
turbulent diffusion of the magnetic field or turbulent
diamagnetism. In many investigations on the kinematic level the
magnetic fluctuations are understood as caused by the action of
the velocity fluctuations on the mean magnetic field. This implies
that they vanish if the mean magnetic field does so. The mean
electromotive force can then be considered as a quantity
determined, apart from the mean velocity, by the velocity
fluctuations and the mean magnetic field.

There are, however, many realistic cases with magnetic
fluctuations which exist independent of the action of the velocity
fluctuations on the mean magnetic field and do not vanish if it
does so. We recall here the numerous investigations showing the
possibility of small--scale dynamos with a zero mean magnetic
field (see, e.g., Kazantsev 1968, Meneguzzi et al. 1981, Zeldovich
et al. 1990, Nordlund et al. 1992, Childress \& Gilbert 1995,
Brandenburg et al. 1996, Rogachevskii \& Kleeorin 1997, Kleeorin
et al. 2002b). As pointed out already by Pouquet et al. (1976) in
such cases the mean electromotive force has in addition to the
contributions mentioned above, which can be ascribed to the
velocity fluctuations, also others due to those magnetic
fluctuations which exist independent of the mean magnetic field.

Several investigations on the mean electromotive force comprising
both kinds of contributions have been carried out so far. We
mention in particular those by Kichatinov 1982 and by Vainshtein
and Kichatinov 1983, in which an isotropic turbulence with a
scale--independent correlation time was assumed. Results
concerning the $ \alpha $-effect, the turbulent magnetic
diffusivity and the turbulent diamagnetism or paramagnetism  for
an originally isotropic turbulence subject to mean rotation have
been derived using a modified second--order correlation
approximation by Kichatinov 1991, R\"{u}diger \& Kichatinov 1993
and  Kichatinov et al. 1994.

In this paper we present an approach to the mean electromotive
force which reproduces such results under more general assumptions
and reveals new ones, which are of particular importance for
astrophysical applications. We exclude mean motions of the fluid
other than a rigid body rotation with some small rotation rate. We
further consider only a weak mean magnetic field so that its
energy density is small compared to the kinetic energy density. In
this way we do not consider quenching effects, i.e. reductions of
the coefficients defining the mean electromotive force with
growing mean magnetic field, which are very important in view of
the nonlinear behavior of dynamos and have been discussed in a
number of papers (see, e.g., Gruzinov \& Diamond 1994, Cattaneo \&
Hughes 1996, Seehafer 1996, Kulsrud 1999, Field et al. 1999,
Rogachevskii \& Kleeorin 2001, Kleeorin et al. 2002a, Blackman \&
Brandenburg 2002). Finally we use a $\tau$-approximation in the
equations describing the deviation of the cross--helicity tensor
from that for zero magnetic field. In contrast to the often used
second-order correlation approximation it does not totally ignore
higher than second--order correlations but considers their
influence in some summary way. The $\tau$--approximation in that
sense applies in the limit of high hydrodynamic Reynolds numbers.

In Section 2 we will explain the concept of
mean--field magnetohydrodynamics for a homogeneous incompressible
fluid. In Section 3 we introduce a Fourier representation of the
velocity and magnetic fluctuations, define correlation tensors and
express the mean electromotive force by the cross--helicity
tensor. After giving an equation for this tensor in Section 4 we
introduce in Section 5 the mentioned $\tau$-approximation, which
leads to closed equations for that part of the cross--helicity
tensor which determines the mean electromotive force. In Section 6
general relations for this mean electromotive force are given, and
in Section 7 more specific relations for the case for small
rotation rates. In Section 8 we restrict ourselves to the limit of
weak mean magnetic fields so that the mean electromotive force can
be considered as linear in the mean magnetic field and specify the
correlation tensors for the ``original" turbulence, that is, the
turbulence for zero mean magnetic field. For zero rotation they
are determined by simple assumptions concerning the deviations
from a homogeneous isotropic turbulence, and the influence of a
slow rotation on the velocity fluctuations is calculated by a
perturbation procedure, again based on a $\tau$--approximation. We
further introduce Kolmogorov-type spectra of the relevant
quantities. On this basis we deliver results for the coefficients
defining the mean electromotive force and discuss them with
special attention to the different effects of velocity and
magnetic field fluctuations. Finally, in Section 9 our results are
compared with results of the kinematic approach in the
second-order correlation approximation, some remarks concerning
their range of validity are made, and some prospects are mentioned
concerning the extension of the approach of this paper to related
questions of mean-field magnetohydrodynamics.

\section*{2. FORMULATION OF THE PROBLEM}

We consider a turbulent motion of an electrically conducting incompressible
fluid in interaction with a magnetic field.
Let us assume that the fluid velocity  $ {\bf U} $ and the
magnetic field $ {\bf B} $ are governed by the equations
\begin{eqnarray}
\frac{\partial {\bf U} }{ \partial t } + ({\bf U } \cdot \bec{\nabla })
{\bf U}  &=& - \frac{\bec{\nabla} P}{\rho}
+ \frac{1}{\mu \rho}  (\bec{\nabla} \times {\bf B }) \times {\bf B}
\nonumber \\
&  & + \; \nu \Delta {\bf U} + 2 {\bf U} \times
{\bf \Omega} + {\bf F} \;,
\label{S1} \\
\frac{\partial {\bf B} }{ \partial t } & = & \bec{\nabla} \times
({\bf U} \times {\bf B}) + \eta {\Delta} {\bf B} \;,
\label{S2} \\
\bec{\nabla} \cdot {\bf U} & = & \bec{\nabla} \cdot {\bf B} = 0 \;,
\label{SS2}
\end{eqnarray}
where $ P $ is a modified pressure and $ {\bf F} $ an external force.
We refer to a rotating frame with  $ {\bf \Omega} $ being
the angular velocity responsible for the Coriolis force.
As usual $ \nu $ and  $ \eta $ are  the kinematic and magnetic viscosities,
$ \rho $ is the mass density and  $ \mu $ is the magnetic
permeability of the fluid, all considered as constants.

We further assume  that there is an averaging procedure which
defines for each quantity $ Q $ an average $ \langle Q \rangle $
and satisfies the Reynolds rules. In the spirit of the mean-field
concept we split the fluid velocity $ {\bf U} $ and the magnetic
field $ {\bf  B} $ according to
\begin{eqnarray}
{\bf U} = \overline{\bf U} +  {\bf u} \;, \quad  {\bf  B} =
\overline{\bf B} + {\bf b} \;,
\label{B1}
\end{eqnarray}
into mean parts, $ \overline{\bf U} = \langle {\bf U} \rangle $
and $ \overline{\bf B} = \langle {\bf B} \rangle ,$ and
fluctuations, $ {\bf u} $ and $ {\bf b} .$ Analogously we split $
P $ into $ \overline{P} $ and $ p ,$ and $ {\bf F} $ into $
\overline{\bf F} $ and $ {\bf f} ,$ etc. The mean fields $
\overline{\bf U} $ and $ \overline{\bf B} $ satisfy the equations
\begin{eqnarray}
\frac{\partial \overline{\bf U} }{ \partial t } + (\overline{\bf U }
\cdot \bec{\bf  \nabla }) \overline{\bf U} &=& - \frac{\bec{\nabla}
\overline{P} }{ \rho } + \frac{1}{\mu \rho}
(\bec{\nabla} \times \overline{\bf B }) \times \overline{\bf B}
+ \nu \Delta \overline{\bf U} + 2 \overline{\bf U} \times {\bf \Omega}
+ \bec{\cal F} + \overline{\bf F} \;,
\label{S3} \\
\frac{\partial \overline{\bf B} }{ \partial t } & = & \bec{\nabla} \times
(\overline{\bf U} \times \overline{\bf B} + \bec{\cal E}) + \eta \Delta
\overline{\bf B}   \;,
\label{S4} \\
\bec{\nabla} \cdot \overline{\bf U} & = & \bec{\nabla} \cdot
\overline{\bf B} = 0 \;,
\label{SS4}
\end{eqnarray}
with $  \bec{\cal F} $ and $ \bec{\cal E} $ being a pondermotive and
an electromotive force due to fluctuations,
\begin{eqnarray}
\bec{\cal F} & = & -  \langle ({\bf u} \cdot \bec{\bf  \nabla }) {\bf u}
\rangle
+ \frac{1}{\mu \rho} \langle (\bec{\nabla}
\times {\bf b}) \times {\bf b} \rangle \;,
\label{B2} \\
\bec{\cal E} & = & \langle {\bf u} \times {\bf b} \rangle \; .
\label{B3}
\end{eqnarray}
For the determination of $  {\cal F} $ and $ {\cal E} $
we need information on the fluctuations $ {\bf u} $ and $ {\bf b}. $
These have to obey the equations
\begin{eqnarray}
\frac{\partial {\bf u} }{ \partial t } & = & - (\overline{\bf U}
\cdot \bec{\bf  \nabla }) {\bf u } - ( {\bf u }\cdot \bec{\bf
\nabla }) \overline{\bf U} - \frac{\bec{\bf \nabla}p }{ \rho } +
\frac{1}{\mu \rho} (\bec{\nabla} \times \overline{\bf B }) \times
{\bf b}
\nonumber\\
& & + (\bec{\nabla} \times {\bf b} ) \times \overline{\bf B}] +
{\bf T} + \nu \Delta {\bf u} + 2 {\bf u} \times {\bf \Omega} +
{\bf f} \;,
\label{Q4}\\
\frac{\partial {\bf b} }{ \partial t } & = & \bec{\nabla} \times
({\bf u} \times \overline{\bf B} + \overline{\bf U} \times {\bf b})
+ \eta \Delta {\bf b} + {\bf G} \;,
\label{Q5}
\end{eqnarray}
where $ {\bf T} $ and $ {\bf G} $ summarize terms nonlinear in
$ {\bf u} $ and $ {\bf b} ,$
\begin{eqnarray}
{\bf T} & = & \langle ({\bf u }\cdot \bec{\nabla}){\bf u} \rangle
- ({\bf u} \cdot \bec{\nabla}){\bf u} + {1 \over \mu \rho }
[\langle {\bf b} \times ( \bec{\nabla} \times {\bf b}) \rangle -
{\bf b} \times (\bec{\nabla} \times {\bf b} ) ]  \;,
\label{B4} \\
{\bf G} & = & \bec{\nabla} \times ({\bf u} \times {\bf b} - \langle {\bf
u} \times {\bf b} \rangle ) \; .
\label{B5}
\end{eqnarray}

\section*{3. FOURIER REPRESENTATION, CORRELATION TENSORS}

Let us now represent quantities like $ {\bf u} $ and $ {\bf b} ,$
or $ \overline{\bf U} $ and $ \overline{\bf B} ,$ by Fourier
integrals defined according to
\begin{eqnarray}
Q({\bf x},t) = \int \hat Q({\bf k},t) \exp (i {\bf k} \cdot {\bf
x}) \,d^{3} k \; . \label{L70}
\end{eqnarray}
We may rewrite the equations (\ref{Q4}) and (\ref{Q5}) for $ {\bf u} $
and $ {\bf b} $ into equations for their components $ u_i $ and
$ b_i $ with respect to a Cartesian coordinate system, derive equations for
their Fourier transforms $ \hat u_i $ and $ \hat b_i $
and subject the equation for $ \hat u_i $ to a projection
operator $ P_{ij} =  \delta _{ij} - k_i  k_j / k^2 ,$ with
$ \delta_{ij} $ being the Kronecker tensor, in order to eliminate
the pressure $ p .$ In this way we obtain
\begin{eqnarray}
\partial \hat u_i({\bf k}) / \partial t & = & - P_{ij} [\hat
L_{j}({\bf u,\overline{U}; k}) + \hat S_{j}({\bf u,\overline{U}; k})] +
(2 P_{ij} - \delta_{ij}) \hat L_{j}({\bf b,\overline{B}; k}) / \mu \rho
\nonumber \\
& & + \hat S_{i}({\bf b,\overline{B}; k}) / \mu \rho + D_{ij}({\bf
k})\hat u_{j}({\bf k}) - \nu k^2 \hat u_i({\bf k}) - P_{ij}(\hat
T_j({\bf k}) - \hat f_{j}({\bf k})) / \rho \;,
\label{Q1}\\
\partial \hat b_i({\bf k},t) / \partial t & = & \hat S_{i}({\bf
u,\overline{B}; k})
- \hat S_{i}({\bf b,\overline{U}; k}) - \hat L_{i}({\bf u,\overline{B}; k})
+ \hat L_{i}({\bf b,\overline{U}; k})
\nonumber \\
& & - \eta k^2 \hat b_i({\bf k}) + \hat G_i({\bf k}) \;,
\label{Q2}
\end{eqnarray}
where
\begin{eqnarray}
\hat L_{i}({\bf a,A; k}) &=& i \int a_j ({\bf k}-{\bf K}) K_{j} A_i ({\bf
K}) \,d^{3} K,
\label{B6} \\
\hat S_{i}({\bf a,A; k}) &=& i k_{j} \int a_i ({\bf k}-{\bf K}) A_j ({\bf
K}) \,d^{3} K \;,
\label{B7} \\
D_{ij}({\bf k}) &=& 2 \varepsilon_{ijm} k_{m} ({\bf k} \cdot {\bf \Omega}) /
k^{2} \;,
\label{B8}
\end{eqnarray}
and $ \varepsilon_{ijk} $ is the Levi--Civita tensor. For the sake
of simplicity the argument $ t $ is dropped everywhere. Concerning
the derivation of (\ref{B8}) we refer to (\ref{X1}).

We will use these equations for calculating the two-point
correlation functions.  Let us consider, e.g., the correlation
tensor $ \langle v_i ({\bf x}_1) w_j ({\bf  x}_2) \rangle $ of two
vector fields  $ {\bf v} $ and $ {\bf w} ,$ where  $ {\bf x}_1 $
and $ {\bf x}_2 $ denote two points in space but both fields are
taken at the same time. Using the definition  (\ref{L70}) of the
Fourier transformation and following a pattern introduced by
Roberts and Soward 1975 we write
\begin{eqnarray}
\langle v_i ({\bf x}_1 ) w_j ({\bf  x}_2) \rangle &=& \int \int
\langle \hat v_i  ({\bf  k}_1) \hat w_j ({\bf k}_2) \rangle
\exp{i( {\bf  k}_1 \cdot {\bf x}_1 + {\bf  k}_2 \cdot {\bf x}_2)}
\,d^{3} k_1 \, d^{3} k_2
\nonumber \\
&=& \int \tilde \varphi_{ij}( {\bf r, K} ) \exp{(i {\bf  K \cdot
R}) } \,d^{3} K
\nonumber \\
&=& \int \varphi_{ij}( {\bf k, R} ) \exp{(i {\bf k \cdot r}) }
\,d^{3} k \;, \label{B11}
\end{eqnarray}
where
\begin{eqnarray}
\tilde \varphi_{ij}({\bf  r ,  K} ) &=& \int \langle \hat v_i
({\bf k} + {\bf K} / 2 ) \hat w_j( -{\bf k} + {\bf  K}  / 2 )
\rangle \exp{(i {\bf k \cdot r}) } \,d^{3} k \;,
\label{B12} \\
\varphi_{ij}({\bf k, R} ) &=& \int \langle \hat v_i ({\bf k} +
{\bf  K} / 2 ) \hat w_j( -{\bf k} + {\bf  K}  / 2 ) \rangle
\exp{(i {\bf K \cdot R}) } \,d^{3} K\;, \label{B13}
\end{eqnarray}
and $ {\bf R} = ( {\bf x}_1 +  {\bf x}_2) / 2  , \quad
{\bf r} = {\bf x}_1 - {\bf x}_2, \quad {\bf K} = {\bf k}_1 + {\bf k}_2,
\quad {\bf k} = ( {\bf k}_1 - {\bf k}_2) / 2 .$
We relate later $ {\bf r} $ and $ {\bf k} $ to  small scales and
$ {\bf R} $ and $ {\bf K} $ to large scales in the physical space.

In the following we consider in particular the correlation tensors
for the velocity and magnetic fluctuations, $ \langle u_i ({\bf
x}_1) u_j ({\bf x}_2) \rangle $ and $ \langle b_i ({\bf x}_1) b_j
({\bf x}_2) \rangle ,$ and the cross-helicity tensor, $ \langle
u_i ({\bf x}_1) b_j ({\bf x}_2) \rangle ,$ and we use the
definitions
\begin{eqnarray}
v_{ij}({\bf k, R}) &=& \Phi ( \hat u_i, \hat u_j; {\bf k}, {\bf R} ) \; ,
\quad m_{ij}({\bf k, R}) = \Phi ( \hat b_i, \hat b_j; {\bf k}, {\bf R} ) /
\mu \rho \; ,
\label{B14} \\
\chi_{ij}({\bf k, R}) &=& \Phi ( \hat u_i, \hat b_j; {\bf k}, {\bf R} ) \; ,
\label{B15}
\end{eqnarray}
where
\begin{equation}
\Phi (v, w; {\bf k}, {\bf R} ) = \int \langle v({\bf k} + {\bf  K}
/ 2) w(-{\bf k} + {\bf  K}  / 2 ) \rangle \exp{(i {\bf K \cdot R})
} \,d^{3} K \; . \label{B16}
\end{equation}
The definition of $v_{ij}$ implies
\begin{equation}
v_{ij}({\bf k, R}) = v_{ji}({\bf - k, R})
\label{Y1}
\end{equation}
and
\begin{equation}
v_{ij}({\bf k}, {\bf R}) k_i = \frac{i}{2} \nabla_i v_{ij}({\bf
k}, {\bf R}) \; , \quad v_{ij}({\bf k}, {\bf R}) k_j = -
\frac{i}{2} \nabla_j v_{ij}({\bf k}, {\bf R}) \; . \label{Y2}
\end{equation}
Here and in the following ${\bec \nabla}$ stands for $\partial /
\partial {\bf R}$. A relation analogous to (\ref{Y1}) applies to
$m_{ij}$, too, and relations analogous to (\ref{Y2}) apply to
$m_{ij}$ and even to $\chi_{ij}$.

If we know $ v_{ij} ,$ $ m_{ij} $ and $ \chi_{ij} $ we may calculate
the pondermotive force $ \bec{\cal F} $ and the electromotive force
$ \bec{\cal E} .$ In this paper, however, we will focus attention
on $ \bec{\cal E} $ only. Since
\begin{eqnarray}
{\cal E}_{i} = \varepsilon_{ijk} \int \chi_{jk}({\bf  k , R}) \,d^{3} k
\label{B60}
\end{eqnarray}
we first deal with the cross-correlation tensor $ \chi_{ij} .$

\section*{4. THE EQUATION FOR THE CROSS-HELICITY TENSOR}

According to the definition (\ref{B16}) of $ \chi_{ij} $
its time derivative is given by
\begin{equation}
{\partial \chi_{ij}({\bf k, R}) \over \partial t}
= \Phi ( \frac{\partial \hat u_i}{\partial t}, \hat b_j; {\bf k}, {\bf R} )
+ \Phi ( \hat u_i, \frac{\partial \hat b_j}{\partial t}; {\bf k}, {\bf R} )
\; .
\label{QQ3}
\end{equation}
We evaluate this using equations (\ref{Q1}) and (\ref{Q2}) and
neglecting all contributions containing higher than first order
terms in the operator $\bec{\nabla}$, independent on whether it
acts on $ v_{ij} ,$ $ m_{ij} ,$ $ \chi_{ij} ,$ $ \overline{\bf U}
$ or $ \overline{\bf B} .$ With manipulations explained in
Appendix B we obtain
\begin{eqnarray}
{\partial \chi_{ij} \over \partial t} +
L_{ijklm} {\partial \chi_{kl} \over \partial k_{m}}
+ M_{ijklm} \nabla_{m} \chi_{kl}
+ N_{ijkl} \chi_{kl} + C_{ij} = I_{ij} \;,
\label{Q3}
\end{eqnarray}
where
\begin{eqnarray}
L_{ijklm} &=& - \delta_{ik} \delta_{jl} \overline{U}_{p,m} k_{p} \;,
\label{B17} \\
M_{ijklm} &=& - i ( D_{ik} \delta_{jl} k_{m} -  \varepsilon_{ikm}
\delta_{jl} ({\bf k} \cdot {\bf \Omega}) - \varepsilon_{ikp}
k_{p} \delta_{jl} \Omega_{m} ) / k^{2}
\nonumber \\
& & + i (\eta - \nu) \delta_{ik} \delta_{jl} k_{m} \;,
\label{B18} \\
N_{ijkl} &=& \overline{U}_{i,k} \delta_{jl} - \delta_{ik}
\overline{U}_{j,l}  - 2 k_{ip} \overline{U}_{p,k}
\delta_{jl} - D_{ik} \delta_{jl} + (\nu + \eta) k^{2} \delta_{ik}
\delta_{jl} \;,
\label{B19} \\
C_{ij} &=& \Phi(P_{ik}(\hat T_k - \hat f_k ), \hat b_j) - \Phi(\hat u_i,
\hat G_j) \;,
\label{B20} \\
I_{ij} &=& - i({\bf k } \cdot \overline{\bf B}) (v_{ij} - m_{ij} ) +
\frac{1}{2} \overline{B}_{p} \nabla_{p} (v_{ij} + m_{ij})
\nonumber\\
& & - \overline{B}_{j,p} v_{ip} + \overline{B}_{i,p} m_{pj} - 2
\frac{k_i k_p}{k^2} \overline{B}_{p,q} m_{qj} - \overline{B}_{p,q}
k_{p} (v_{ijq} + m_{ijq}) \;, \label{B21}
\end{eqnarray}
and $ \overline{B}_{i,j} = \nabla_{j} \overline{B}_{i} ,$
$ \overline{U}_{i,j} = \nabla_{j} \overline{U}_{i} ,$
furthermore $ k_{ij} = k_{i} k_{j} / k^{2} ,$
$ v_{ijq} = \frac{1}{2} \partial v_{ij} / \partial k_{q} $ and
$ m_{ijq} = \frac{1}{2} \partial m_{ij} / \partial k_{q} .$

\section*{5. THE $\tau$-APPROXIMATION}

When dealing with equation (\ref{Q3}) for $ \chi_{ij} $ we are
confronted with the difficulty that we know nothing about the term
$ C_{ij} $ which stands for couplings of $ \chi_{ij} $ with
higher-order correlation tensors. We assume now that we know the
solution $ \chi_{ij} $ of equation (\ref{Q3}) for vanishing mean
magnetic field, that is $ I_{ij} = 0 ,$ denote it by  $
\chi_{ij}^{(0)} ,$ and the corresponding $ C_{ij} $ by $
C_{ij}^{(0)} .$ For the case with non-vanishing mean magnetic
field we put then $ \chi_{ij} = \chi_{ij}^{(0)} + \chi_{ij}^{(B)}
$ and $ C_{ij} = C_{ij}^{(0)} + C_{ij}^{(B)} .$ Then we have
\begin{eqnarray}
{\partial \chi_{ij}^{(B)} \over \partial t} +
L_{ijklm} {\partial \chi_{kl}^{(B)} \over \partial k_{m}}
+ M_{ijklm} \nabla_{m} \chi_{kl}^{(B)}
+ N_{ijkl} \chi_{kl}^{(B)} + C_{ij}^{(B)} = I_{ij} \; .
\label{Q3Q}
\end{eqnarray}
At this level we introduce the $\tau$-approximation
\begin{eqnarray}
C_{ij}^{(B)} = \chi_{ij}^{(B)} / \tau (k) \;,
\label{Q7}
\end{eqnarray}
with some relaxation time $ \tau (k) $ (see, e.g., Orszag 1970,
Monin \& Yaglom 1975, Pouquet et al. 1976, McComb 1990, Kleeorin
et al. 1990, 1996). This approximation applies for large hydrodynamic
Reynolds numbers, i.e. for fully developed turbulence. In this case the
relaxation time $ \tau (k) $ is determined by the correlation
time of the turbulent velocity field only.
The limit $ \tau \to \infty $ corresponds to cancelling all
higher-order correlations, that is, to some kind of second-order
correlation approximation (see also Section 9). Note that we do
not introduce any closure assumption for the original turbulence;
equation (\ref{Q7}) concerns only deviations from the original
turbulence.

\section*{6. RELATIONS FOR THE MEAN ELECTROMOTIVE FORCE}

Returning now to the electromotive force $ \bec{\cal E} $ we first
put in the above sense $ \bec{\cal E} = \bec{\cal E}^{(0)} +
\bec{\cal E}^{(B)} .$ We assume however that $ \bec{\cal E}^{(0)}
,$  if not at all equal to zero, becomes unimportant in comparison
to $ \bec{\cal E}^{(B)} $ when $ \overline{\bf B} $ grows and put
therefore $ \bec{\cal E}^{(0)} = 0 .$ Thus the electromotive force
$ \bec{\cal E} $ is given by
\begin{eqnarray}
{\cal E}_{i} = \varepsilon_{ijk} \int \chi_{jk}^{(B)}({\bf  k , R})
\,d^{3} k \; .
\label{Q7Q}
\end{eqnarray}

Let us assume that the time variations of $ I_{ij} ,$ that is, of
$ \overline{B}_i ,$ $ \overline{B}_{i,j} ,$ $ v_{ij} ,$ and $
m_{ij} ,$ are sufficiently weak, so that $ \chi_{ij}^{(B)} ,$
apart from some initial time, can be considered as a solution of
the steady version of equation (\ref{Q3Q}), that is, with $
\partial \chi_{ij}^{(B)} /  \partial t $  being neglected. Thus
the dependence of $ \chi_{ij}^{(B)} $ on $ \overline{B}_i $ and $
\overline{B}_{i,j} $ is linear and homogeneous and, in addition,
instantaneous. So we may conclude that
\begin{equation}
{\cal E}_{i} = a_{ij} \overline{B}_{j} + b_{ijk} \overline{B}_{j,k}
\label{Q52}
\end{equation}
with tensors $ a_{ij} $ and $ b_{ijk} $ determined by $ v_{ij} $
and $ m_{ij} .$

This relation can be rewritten in another form, which might be
more suitable for some discussions. To derive it we split $ a_{ij}
$ in a symmetric and an antisymmetric part and express the latter
by a vector. We further split the gradient tensor of $
\overline{\bf B} $ into a symmetric  part $ \bec{\partial
\overline{\bf B}} $, defined by $ (\partial \overline{B})_{ij} =
\frac{1}{2}(\overline{B}_{i,j} + \overline{B}_{j,i}) $, and an
antisymmetric one, which can be expressed by $ \bec{\bf \nabla}
{\bf \times} \overline{\bf B} $. We have then $ \overline{B}_{i,j}
= (\partial \overline{B})_{ij} - \frac{1}{2} \varepsilon_{ijk}
(\bec{\bf \nabla} {\bf \times} \overline{\bf B})_{k} $. Finally we
express the tensorial coefficient occurring then with $ \bec{\bf
\nabla} {\bf \times} \overline{\bf B} $ again by a symmetric
tensor and a vector. In this way we arrive at
\begin{eqnarray}
\bec{\cal E} = - \bec{\alpha} \overline{\bf B} - \bec{\gamma}
{\bf \times} \overline{\bf B} - \bec{\beta} (\bec{\nabla}
{\bf \times} \overline{\bf B}) - \bec{\delta} {\bf \times}
(\bec{\nabla} {\bf \times} \overline{\bf B}) - \bec{\kappa}
\bec{\partial \overline{\bf B}} \;
\label{L1}
\end{eqnarray}
(see R\"{a}dler 1980, 1983), where $ \bec{\alpha} $ and $
\bec{\beta} $ are symmetric tensors of the second rank, $
\bec{\gamma} $ and $ \bec{\delta} $ vectors, and $ \bec{\kappa} $
is a tensor of the third rank; $ \bec{\kappa} $ can be considered
to be symmetric in the indices connecting it with $ \bec{\partial
\overline{\bf B}} ,$ and contributions can be dropped which would
produce $ \bec{\nabla} \cdot \overline{\bf B}$. We have
\begin{eqnarray}
\alpha_{ij} &=& - \frac{1}{2} (a_{ij} + a_{ji}) \; ,
\quad \beta_{ij} = \frac{1}{4} (\varepsilon_{ikl} b_{jkl} +
\varepsilon_{jkl} b_{ikl}) \; ,
\label{La} \\
\gamma_{i} &=& \frac{1}{2} \varepsilon_{ijk} a_{jk} \; ,
\quad \delta_{i} = \frac{1}{4} (b_{jji} - b_{jij}) \; ,
\quad \kappa_{ijk} = - \frac{1}{2} (b_{ijk} + b_{ikj}) \; .
\label{Lb}
\end{eqnarray}

For the sake of simplicity we restrict ourselves on the case $
\overline{U}_i =$ const. Thus $ \chi_{ij}^{(B)} $ is governed by
the steady version of equation (\ref{Q3Q}) with $L_{ijklm}=0$ and
$N_{ijkl}= - D_{ik} \delta_{jl} + (\nu + \eta) k^2 \delta_{ik}
\delta_{jl}$. Using (\ref{Q7}) the equation for $ \chi_{ij}^{(B)}
$ can be written in the form
\begin{eqnarray}
\tilde D_{ik} \chi_{kj}^{(B)} + M_{ijklm} \nabla_{m} \chi_{kl}^{(B)}
\tau_{\ast} = I_{ij} \tau_{\ast} \;,
\label{S12}
\end{eqnarray}
where
\begin{eqnarray}
\tilde D_{ij} = \delta_{ij} -  D_{ij} \tau_{\ast} \;, \quad
\tau_{\ast}^{-1} = \tau^{-1}  + (\nu + \eta) k^2 \; .
\label{B22}
\end{eqnarray}
In deriving equation (\ref{Q3}) we have neglected all
contributions to $ I_{ij} $ of  higher than first order in $
\bec{\nabla} .$ In the same sense we may replace $ \nabla_{m}
\chi_{kl}^{(B)} $ in equations (\ref{Q3}), (\ref{Q3Q}) and
(\ref{S12}) by $ \nabla_{m} \chi_{kl}^{o(B)} ,$ where $
\chi_{ij}^{o(B)} $ summarizes the contributions to $
\chi_{ij}^{(B)} $ which are of zero order in $ \bec{\nabla} .$ We
have then
\begin{eqnarray}
\tilde D_{ik} \chi_{kj}^{o(B)} =  I_{ij}^{o} \tau_{\ast} \;,
\quad I_{ij}^{o} = - i(\overline{\bf B} \cdot {\bf k }) (v_{ij}
- m_{ij}) \;,
\label{S14}
\end{eqnarray}
and consequently
\begin{eqnarray}
\chi_{ij}^{o(B)} =  \tilde D_{il}^{-1} I_{lj}^{o} \tau_{\ast} \;,
\label{S15}
\end{eqnarray}
where $ \tilde {\bf D}^{-1} $ is the inverse of $ \tilde {\bf D}
,$ satisfying $ \tilde D_{ik}^{-1} \tilde D_{kj} = \delta_{ij} ,$
that is,
\begin{eqnarray}
\tilde D_{ij}^{-1} = (1 + \omega^2 k^2)^{-1} (\delta_{ij} + \omega
\varepsilon_{ijk} k_k + \omega^2 k_i k_j) \;,
\label{B23}
\end{eqnarray}
where $ \omega = 2 \tau_{\ast} ({\bf \Omega} \cdot {\bf k }) / k^2 .$
Proceeding as described above we finally obtain the solution of
equation (\ref{S12}) in the form
\begin{eqnarray}
\chi_{ij}^{(B)} = \tilde D_{ik}^{-1} I_{kj} \tau_{\ast} -
\tilde D_{ip}^{-1} M_{pjklm} \tilde D_{kq}^{-1} \nabla_{m} I_{ql}^{o}
\tau_{\ast}^2 \; .
\label{S16}
\end{eqnarray}

\section*{7. THE MEAN ELECTROMOTIVE FORCE AT SLOW ROTATION}

We split the electromotive force $ \bec{\cal E}^{(B)} $ according to
\begin{eqnarray}
\bec{\cal E}^{(B)} = \bec{\cal E}^{(B0)} + \bec{\cal E}^{(B\Omega)}
\label{S17}
\end{eqnarray}
into parts $ \bec{\cal E}^{(B0)} $ and $ \bec{\cal E}^{(B\Omega)} $,
the first of which does not depend on the rotation while the second one does
but vanishes with vanishing rotation.
For $ {\cal E}_{i}^{(B0)} $ we have simply
\begin{eqnarray}
{\cal E}_{i}^{(B0)} = \varepsilon_{ijk} \int I_{jk}({\bf k}) \tau_{\ast}(k)
\,d^3 k \; .
\label{B24}
\end{eqnarray}
As for $ \bec{\cal E}^{(B\Omega)} $ we restrict ourselves to the
case of slow rotation. More precisely, we neglect terms of third
and higher order in $\Omega \tau_{\ast}$ in comparison to unity.
So we find
\begin{eqnarray}
{\cal E}_{i}^{(B\Omega)} &=&
\int \biggl( 2 ({\bf k} \cdot {\bf \Omega}) (k_i I_{jj} - k_j I_{ij})
\nonumber \\
& & + i ( ( 2 \frac{({\bf k} \cdot {\bf \Omega})}{k^2} ({\bf k}
\cdot {\bf \nabla}) - ({\bf \Omega} \cdot {\bf \nabla}) ) k_i
I_{jj}^{(o)} - ({\bf k} \cdot {\bf \Omega}) (\nabla_i I_{jj}^{(o)}
- \nabla_j I_{ij}) )
\nonumber \\
& & - 4 ({\bf k} \cdot {\bf \Omega})^2 (\varepsilon_{ijk} +
\varepsilon_{ikl} \frac{k_l k_j}{k^2}) I_{jk} \tau_{\ast}
\nonumber \\
& & - 2 i ({\bf k} \cdot {\bf \Omega}) (2(\frac{({\bf k} \cdot
{\bf \Omega})}{k^2} ({\bf k} \cdot {\bf \nabla}) - ({\bf \Omega}
\cdot {\bf \nabla})) \varepsilon_{ijk} I_{jk}^{(o)}
\nonumber \\
& & + \frac{({\bf k} \cdot {\bf \Omega})}{k^2} \varepsilon_{ijk}
k_j \nabla_{l} I_{lk}^{(o)} ) \tau_{\ast} \biggr)
\frac{\tau_{\ast}^2}{k^2} d^3 k \; . \label{B25}
\end{eqnarray}
In the integrand we have dropped terms of the structures $\nabla_j
k_k I_{kl}^{(o)}$ or $\nabla_j k_k I_{lk}^{(o)}$, which would, by
reason connected with (\ref{Y2}), lead to contributions of the
second order in $\nabla$. Moreover, for the sake of simplicity in
both (\ref{B24}) and (\ref{B25}) we ignored terms containing the
factor $(\eta - \nu)k^2 \tau_{\ast}$. We will later point out the
consequences of that for our final result.

In relation (\ref{Q52}) for $ \bec{\cal E} $ we split $ a_{ij} $
and $ b_{ijk} $ into parts corresponding to their dependence on $
v_{ij} $ or $ m_{ij} $, e.g.,
\begin{equation}
a_{ij} = a_{ij}^{(v)} + a_{ij}^{(m)} \;,
\label{L2} \\
\end{equation}
and each of them are splitted  into one which is independent of $
\Omega $ and a remaining one, e.g.,
\begin{equation}
a_{ij}^{(v)} = a_{ij}^{(v0)} + a_{ij}^{(v\Omega)} \;,
\quad a_{ij}^{(m)} = a_{ij}^{(m0)} + a_{ij}^{(m\Omega)} \; .
\label{L10}
\end{equation}
We note that the splitting of $\bec{\cal E}^{(B)}$ according to
(\ref{S17}) and the splittings of $a_{ij}^{(v)}$, $a_{ij}^{(m)}$,
$\cdots$ according to (\ref{L10}) refer only to the dependencies
on ${\bf \Omega}$ which occur explicitly in this stage of our
derivations. In general $v_{ij}$ will depend on ${\bf \Omega}$,
too, and then $\bec{\cal E}^{(B0)}$ and $a_{ij}^{(v0)}$,
$a_{ij}^{(m0)}$, $\cdots$ will have also contributions with ${\bf
\Omega}$, and the dependencies of $\bec{\cal E}^{(B \Omega)}$ and
$a_{ij}^{(v \Omega)}$, $a_{ij}^{(m \Omega)}$, $\cdots$ on ${\bf
\Omega}$ will be more complex.

A straightforward calculation yields
\begin{eqnarray}
a_{ij}^{(v0)} &=& \varepsilon_{ilm} v_{lmj}^{(1)} \;,
\label{B26} \\
a_{ij}^{(m0)} &=& - \varepsilon_{ilm} m_{lmj}^{(1)} \;,
\label{BB26} \\
b_{ijk}^{(v0)} &=& \varepsilon_{ijl} v_{lk}^{(1)} \;,
\label{B27} \\
b_{ijk}^{(m0)} &=& \varepsilon_{ijl} m_{lk}^{(1)}
+ 2 \varepsilon_{ilm} m_{kljm}^{(1)} \;,
\label{BB27} \\
a_{ij}^{(v\Omega)} &=& \bigl( 2 v_{ppijl}^{(2)}
-  \nabla_i v_{ppjl}^{(2)} + \nabla_j  v_{ppil}^{(2)}
+ 2 \nabla_p ( v_{qqijlp}^{(2)} + v_{ipjl}^{(2)})
- \nabla_l v_{ppij}^{(2)} \bigr) \Omega_l
\nonumber \\
& & - 4 \varepsilon_{ipq} v_{pqjlm}^{(3)} \Omega_l \Omega_m \; ,
\label{B28} \\
a_{ij}^{(m\Omega)} &=& \bigl( - 2 m_{ppijl}^{(2)}
+ \nabla_i m_{ppjl}^{(2)} + \nabla_j  m_{ppil}^{(2)}
- 2 \nabla_p ( m_{qqijlp}^{(2)} + m_{ipjl}^{(2)})
+ \nabla_l m_{ppij}^{(2)} \bigr) \Omega_l
\nonumber \\
& & + 4 \varepsilon_{ipq} m_{pqjlm}^{(3)} \Omega_{l} \Omega_{m} \;
,
\label{BB28} \\
b_{ijk}^{(v\Omega)} &=& \bigl( 2 v_{ikjl}^{(2)} - 2 v_{jkil}^{(2)}
+ v_{ppijkl}^{\prime} \bigr) \Omega_{l}
\nonumber \\
& & - 4 \varepsilon_{ijp} v_{pklm}^{(3)} \Omega_{l} \Omega_{m} \;,
\label{B29} \\
b_{ijk}^{(m\Omega)} &=& \bigl( - 2 m_{ikjl}^{(2)} + 2 m_{jkil}^{(2)}
+ 2 m_{ppij}^{(2)} \delta_{kl} + 2 m_{ppjl}^{(2)} \delta_{ik}
- 4 m_{ppijkl}^{(2)} + m_{ppijkl}^{\prime} \bigr) \Omega_{l}
\nonumber \\
& & - 4 ( \varepsilon_{ijp} m_{kplm}^{(3)} + 2 \varepsilon_{ipq}
m_{kpjlmq}^{(3)}) \Omega_{l} \Omega_{m} \; . \label{B30}
\end{eqnarray}
Here we used the definitions
\begin{eqnarray}
v_{ijk...p}^{(\mu)} &=& \int (-ik)^{\lambda} v_{ij}({\bf k})  k_{k...p}
\tau_{\ast}^{\mu} \,d^3 k \;,
\label{B31} \\
v_{ijklmn}^{\prime} &=& \int v_{ij}({\bf k}) k_{klmn}
(d\tau_{\ast}^{2}/dk) \, k \,d^3 k \;,
\label{B32}
\end{eqnarray}
where $ \lambda = 0 $ or $ \lambda = 1$ for even or odd $\nu$,
respectively, and $ k_{k...p} = k_k ... k_p / k^\nu ,$ with $ \nu
$ being the rank of this last tensor, and analogous definitions
for $ m_{ijk...p}^{(\mu)} ,$ and $ m_{ijklmn}^{\prime}$ with $
v_{ij} $ replaced by $ m_{ij} .$ With the help of (\ref{Y1}) we
find that the $v_{ijk...p}^{(\mu)}$ with $\lambda = 0$ are
symmetric and those with $\lambda = 1$ antisymmetric in  $i$ and
$j$ and moreover in both cases symmetric in every pair of the
remaining indices, and that $v_{ijklmn}^{\prime}$ is symmetric in
$i$ and $j$ and again in in every pair of the remaining indices.
The same applies to $m_{ijk...p}^{(\mu)}$ and to
$m_{ijklmn}^{\prime}$. Furthermore in the case $\lambda = 0$ we
may conclude from (\ref{Y2}) that $v_{pij...np}^{(\mu)} = - (1/2)
\nabla_p {\tilde v}_{pij...n}^{(\mu)}$ and $v_{ipj...np}^{(\mu)} =
(1/2) \nabla_p {\tilde v}_{ipj...n}^{(\mu)}$, where the ${\tilde
v}_{ij...p}^{(\mu)}$ (which will never explicitly occur in the
following) are defined like the $v_{ij...p}^{(\mu)}$ but with
$k_{ij...p} k^{-2}$ in the integrand instead of $k_{ij...p}$. In
the case $\lambda = 1$ we have simply $v_{pij...np}^{(\mu)} =
(1/2) \nabla_p v_{pij...n}^{(\mu)}$ and $v_{ipj...np}^{(\mu)} = -
(1/2) \nabla_p v_{ipj...n}^{(\mu)}$. All these statements apply
again to $m_{pij...np}^{(\mu)}$ and $m_{ipj...np}^{(\mu)}$, too.
We have used these properties of the $v_{ijk...p}^{(\mu)}$,
$m_{ijk...p}^{(\mu)}, \cdots$. in deriving
(\ref{B26})--(\ref{B30}), and we have ignored  all terms which
would result in contributions to $\bec{\cal E}$ of higher than
first order in ${\bf \nabla}$. Likewise we have cancelled
contributions to $b_{ijk}$ proportional to $\delta_{jk}$, which
can not contribute to $\bec{\cal E}$ since ${\bf \nabla} \cdot
\overline{{\bf B}} = 0$.

Let us consider the representation (\ref{L1}) for $ \bec{\cal E} $
and split each of the quantities $ \bec{\alpha} ,$ $ \bec{\gamma} ,$
$ \bec{\beta} ,$ $ \bec{\delta} $ and $ \bec{\kappa} $
after the pattern of (\ref{L2}) and (\ref{L10}) into four parts, e.g.,
\begin{equation}
\alpha_{ij} = \alpha_{ij}^{(v)} + \alpha_{ij}^{(m)} \;,
\label{B38}
\end{equation}
and
\begin{equation}
\alpha_{ij}^{(v)} = \alpha_{ij}^{(v0)} + \alpha_{ij}^{(v\Omega)} \;,
\quad
\alpha_{ij}^{(m)} = \alpha_{ij}^{(m0)} + \alpha_{ij}^{(m\Omega)} \;.
\label{BB38}
\end{equation}
The above remarks concerning the dependencies on ${\bf \Omega}$
apply here analogously. Using (\ref{La})-(\ref{Lb}) and
(\ref{B26})-(\ref{B30}) all these contributions to $ \bec{\alpha}
,$ $ \bec{\gamma} ,$ $ \bec{\beta} ,$ $ \bec{\delta} $ and $
\bec{\kappa} $ can be expressed by the $v_{ijk...p}^{(\mu)}$,
$m_{ijk...p}^{(\mu)}$, $\cdots$. For the sake of simplicity we
give here only the contributions which do not depend on ${\bf
\Omega}$. They read
\begin{eqnarray}
\alpha_{ij}^{(v0)} &=& \frac{1}{2} (\varepsilon_{ilk} v_{klj}^{(1)} +
\varepsilon_{jlk} v_{kli}^{(1)})
\;,
\quad
\alpha_{ij}^{(m0)} = - \frac{1}{2} (\varepsilon_{ilk} m_{klj}^{(1)}
+ \varepsilon_{jlk} m_{kli}^{(1)})
\;,
\label{B39} \\
\beta_{ij} ^{(v0)} &=& \frac{1}{2} (v_{pp}^{(1)} \delta_{ij} - v_{ij}^{(1)})
\;,
\quad
\beta_{ij}^{(m0)} = - \frac{1}{2} (m_{pp}^{(1)} \delta_{ij} - m_{ij}^{(1)}
- 2 m_{ppij}^{(1)})
\;,
\label{B40} \\
\gamma_{i}^{(v0)} &=& \frac{1}{2} {\nabla}_{i} v_{ij}^{(1)}  \;,
\quad \gamma_{i}^{(m0)} = - \frac{1}{2} {\nabla}_{i} m_{ij}^{(1)} \;,
\quad
\delta_{i}^{(v0)} = \delta_{i}^{(m0)} = 0 \;,
\label{B41} \\
\kappa_{ijk}^{(v0)} &=& - \frac{1}{2} (\varepsilon_{ijl}
v_{lk}^{(1)} + \varepsilon_{ikl} v_{lj}^{(1)})
\;,
\label{BB41} \\
\kappa_{ijk}^{(m0)} &=& - \frac{1}{2} (\varepsilon_{ijl}
m_{lk}^{(1)} + \varepsilon_{ikl} m_{lj}^{(1)})
+ \varepsilon_{ilm} ( m_{kmjl}^{(1)} + m_{jmkl}^{(1)})
\; .
\label{B43}
\end{eqnarray}
In order to derive (\ref{B40}) we have used (\ref{X2}).

\section*{8. THE MEAN ELECTROMOTIVE FORCE IN THE LIMIT OF WEAK
MEAN MAGNETIC FIELDS UNDER SPECIFIC ASSUMPTIONS ON THE ORIGINAL
TURBULENCE}

In general the correlation tensors $ v_{ij} $ and $ m_{ij} $
depend, of course, on the mean magnetic field $ \overline{\bf B}
$. As a consequence the mean electromotive force $ \bec{\cal{E}} $
depends in a nonlinear way on $ \overline{\bf B} $. We restrict
ourselves now to the approximation in which $ v_{ij} $ and $
m_{ij} $ are simply replaced by the corresponding tensors $
v^{(0)}_{ij} $ and $ m^{(0)}_{ij} $ for the ``original"
turbulence, that is the turbulence for zero mean magnetic field.
Then, of course, $ \bec{\cal{E}} $ is linear in $ \overline{\bf B}
$. This implies that any quenching effects are excluded.

\subsection*{8.1. Nonrotating turbulence}

We first ignore any influence of a rotation of the fluid on the
turbulence. As for the correlation tensors $ v^{(0)}_{ij} $ and $
m^{(0)}_{ij} $ for the original turbulence we assume that, as long
it is homogeneous, they are essentially determined by the kinetic
and magnetic energy densities, that is by $ \langle {\bf u}^{(0)2}
\rangle $ and $ \langle {\bf b}^{(0)2} \rangle $, and by the
kinematic and current helicities, $ \langle {\bf u}^{(0)} \cdot
(\bec{\nabla}  \times {\bf u}^{(0)}) \rangle $ and $ \langle {\bf
b}^{(0)} \cdot (\bec{\nabla} \times {\bf b}^{(0)}) \rangle $, and
that inhomogeneities are only due to gradients of these four
quantities. With the notations $ {\bf u}^{(0)} $ and $ {\bf
b}^{(0)} $ instead of $ \bf u $ and $ \bf b $ we want to stress
that we are dealing with the original turbulence. Note that $ \bf
b $ is in general non-zero even if $ {\bf b}^{(0)} = \bf{0}$.
Under the assumptions adopted the most general form of the
correlation tensor $ v^{(0)}_{ij}({\bf k},{\bf R}) $ of the
velocity fluctuations of the original turbulence is given by
\begin{eqnarray}
v^{(0)}_{ij}({\bf k},{\bf R}) &=& \frac{1}{8 \pi k^{2}}
\biggl( [ P_{ij}({\bf k}) + \frac{i}{2 k^2}(k_i \nabla_j - k_j \nabla_i) ]
W^{(v)}(k,{\bf R})
\nonumber \\
& & - \frac{1}{2 k^2} [ \varepsilon_{ijk} k_k ( 2 i +
\frac{1}{k^2} ({\bf k} \cdot \bec{\nabla}) ) - \frac{1}{k^2} (k_i
\varepsilon_{jlm} + k_j \varepsilon_{ilm}) k_l \nabla_m ]
\mu^{(v)}(k,{\bf R}) \biggr) \; . \label{L18}
\end{eqnarray}
We note that $ v^{(0)}_{ij} $ satisfies the requirements resulting
from $ \bec{\nabla} \cdot {\bf u}^{(0)} = 0 ,$ and that $
W^{(v)}(k,{\bf R}) $ and $ \mu^{(v)}(k,{\bf R}) $ are spectrum
functions depending on $ {\bf k} $ via $ k $ only and possessing
the properties $\int_0^{\infty} W^{(v)}(k,{\bf R}) \,d k = \langle
{\bf u}^{(0)2} \rangle$ and $\int_0^{\infty} \mu^{(v)}(k,{\bf R})
\,d k = \langle {\bf u}^{(0)} \cdot (\bec{\nabla} \times {\bf
u}^{(0)}) \rangle$. The definition of the correlation tensor $
m_{ij}^{(0)} $ of the magnetic fluctuations of the original
turbulence follows from (\ref{L18}) when replacing $ v^{(0)}_{ij}
,$ $ W^{(v)} $ and $ \mu^{(v)} $ by $ m^{(0)}_{ij} ,$ $ W^{(m)} $
and $ \mu^{(m)} ,$ and we have $\int_0^{\infty} W^{(m)}(k,{\bf R})
\,d k = \langle {\bf b}^{(0)2} \rangle / \mu \rho$ and
$\int_0^{\infty} \mu^{(m)}(k,{\bf R}) \,d k = \langle {\bf
b}^{(0)} \cdot (\bec{\nabla} \times {\bf b}^{(0)}) \rangle / \mu
\rho$.

Since we ignore here any influence of rotation the
$\alpha_{ij}^{(v)}$, $\alpha_{ij}^{(m)}$, $\cdots$ coincide with
the $\alpha_{ij}^{(v0)}$, $\alpha_{ij}^{(m0)}$, $\cdots$ given
with (\ref{B39})--(\ref{B43}). Calculating now the $v_{ijk \cdots
p}^{(1)}$ with $v_{ij}$ replaced by $v_{ij}^{(0)}$ as specified by
(\ref{L18}) we find
\begin{eqnarray}
v_{ij}^{(1)} &=& \frac{1}{3} \delta_{ij} I^{(v1)} + \cdots \; ,
\label{Y11} \\
v_{ijk}^{(1)} &=& \frac{1}{12} ( \delta_{ik} \nabla_j - \delta_{jk}
\nabla_i ) I^{(v1)}
- \frac{1}{6} \varepsilon_{ijk} J^{(v1)} \; ,
\label{Y12} \\
v_{ijkl}^{(1)} &=& \frac{1}{6} (\delta_{ij} \delta_{kl}
- \frac{3}{5} E_{ijkl}^{(4)} ) I^{(v1)} + \cdots \; ,
\label{Y13}
\end{eqnarray}
where
\begin{equation}
I^{(v1)} = \int_0^{\infty} W^{(v)}(k,{\bf R}) \tau_{\ast} \,d k \;,
\quad J^{(v1)} = \int_0^{\infty} \mu^{(v)}(k,{\bf R}) \tau_{\ast} \,d k \;,
\label{Y14}
\end{equation}
and $E_{ijkl}^{(4)}$ is a completely symmetric tensor, which we
define with a view to the following in a more general frame by
\begin{equation}
E_{ijkl \cdots pq}^{(2 \nu)} = \frac{1}{2 \nu -1} ( \delta_{ij}
E_{kl \cdots pq}^{(2 \nu - 2)} + \delta_{ik} E_{jl \cdots pq}^{(2
\nu - 2)} + \cdots + \delta_{iq} E_{jkl \cdots p}^{(2 \nu - 2)} )
\; , \quad E_{ij}^{(2)} = \delta_{ij} \; . \label{Y15}
\end{equation}
The terms in (\ref{Y11}) and (\ref{Y13}) indicated by $\cdots$ are
without interest because they would lead to contributions to ${\bf
\cal{E}}$ which are of second order in ${\bf \nabla}$. For the
$m_{ij}^{(1)}$, $m_{ijk}^{(1)}$ and $m_{ijkl}^{(1)}$ analogous
relations apply with analogously defined $I^{(m1)}$ and
$J^{(m1)}$.

Using now (\ref{B39})--(\ref{B43}) we obtain
\begin{eqnarray}
\alpha_{ij}^{(v)} &=& \frac{1}{3} \delta_{ij} J^{(v1)} \;,
\quad \alpha_{ij}^{(m)} = - \frac{1}{3} \delta_{ij} J^{(m1)} \;,
\label{B44} \\
\beta_{ij}^{(v)} &=& \frac{1}{3} \delta_{ij} I^{(v1)} \;,
\quad \beta_{ij}^{(m)} = 0 \;,
\label{BB44} \\
\bec{\gamma}^{(v)} &=& \frac{1}{6} \bec{\nabla} I^{(v1)}\;,
\quad \bec{\gamma}^{(m)} = - \frac{1}{6} \bec{\nabla} I^{(m1)} \;,
\label{B45} \\
\bec{\delta}^{(v)} &=& \bec{\delta}^{(m)} = 0 \;,
\quad \bec{\kappa}^{(v)} = \bec{\kappa}^{(m)} = {\bf 0} \;.
\label{B46}
\end{eqnarray}

\subsection*{8.2. Rotating Turbulence}

In order to include the effect of a rotation of the fluid we have
first to study how the original turbulence, that is $v_{ij}^{(0)}$
changes with the rotation, that is with ${\bf \Omega}$. We assume
here that in the absence of rotation the turbulence possesses no
helicity. More precisely, we assume that (\ref{L18}) applies with
$\mu^{(v)} = 0$ in the limit of vanishing ${\bf \Omega}$, and we
will now calculate the effect of rotation by a perturbation
procedure up to the second order in ${\bf \Omega}$.

Proceeding as in the calculation of the cross--helicity $\chi_{ij}$ we start
from
\begin{equation}
{\partial v_{ij}({\bf k},{\bf R}) \over \partial t}
= \Phi (\frac{\partial \hat u_i}{\partial t}, \hat u_j; {\bf k}, {\bf R} )
+ \Phi (\hat u_i, \frac{\partial \hat u_j}{\partial t}; {\bf k}, {\bf R} )
\label{L41}
\end{equation}
but use the equation (\ref{Q1}) for $\hat u_i$ with $\overline{\bf B} = {\bf
0}$
and ${\bf T} = {\bf 0}$ so that $v_{ij}$ turns into $v_{ij}^{(0)}$, and put
again
$\overline{\bf U} = {\bf 0}$.
So we arrive at
\begin{eqnarray}
{\partial v_{ij}^{(0)} \over \partial t} + \tilde M_{ijklm}
\nabla_{m} v_{kl}^{(0)} + \tilde N_{ijkl} v_{kl}^{(0)} + 2 \nu k^2
v_{ij}^{(0)} + \tilde C_{ij} = 0 \;, \label{L42}
\end{eqnarray}
where
\begin{eqnarray}
\tilde M_{ijklm} &=& \frac{i}{k^2} \Omega_{p}
( P_{mp}({\bf k}) k_q  + P_{mq}({\bf k}) k_p )
(\varepsilon_{ikq} \delta_{jl} - \varepsilon_{jlq} \delta_{ik}) \;,
\label{L43} \\
\tilde N_{ijkl} &=& - 2 \frac{({\bf k} \cdot {\bf \Omega})}{k^2}
(\varepsilon_{ikp} \delta_{jl} + \varepsilon_{jlp} \delta_{ik})) k_p \;,
\label{L44} \\
\tilde C_{ij} &=& - \Phi (\hat u_i, P_{ik} \hat f_j) - \Phi (P_{ik} \hat
f_k, \hat u_j) \; .
\label{L45}
\end{eqnarray}

We assume now that we know the solution $v_{ij}^{(0)}$ of
(\ref{L42}) for ${\bf \Omega} = {\bf 0}$, denote it by
$v_{ij}^{(00)}$ and the corresponding $\tilde C_{ij}$ by $\tilde
C_{ij}^{(0)}$. For the case with non-vanishing ${\bf \Omega}$ we
put then $ v_{ij}^{(0)} = v_{ij}^{(00)} + v_{ij}^{(0 \Omega)} $
and $ \tilde C_{ij} = \tilde C_{ij}^{(0)} + \tilde
C_{ij}^{(\Omega)} .$ So we find
\begin{eqnarray}
{\partial v_{ij}^{(0 \Omega)} \over \partial t} + \tilde M_{ijklm}
\nabla_{m} v_{kl}^{(0 \Omega)} + \tilde  N_{ijkl} v_{kl}^{(0
\Omega)} + 2 \nu k^2 v_{ij}^{(0 \omega)} + \tilde
C_{ij}^{(\Omega)} = I_{ij}^{(\Omega)} \; , \label{L46}
\end{eqnarray}
where
\begin{equation}
I_{ij}^{(\Omega)} = - ( \tilde M_{ijklm} \nabla_{m}
+ \tilde N_{ijkl} ) v_{kl}^{(00)} \, .
\label{LL46}
\end{equation}

At this level we introduce again a $\tau$-approximation,
\begin{equation}
\tilde C_{ij}^{(\Omega)} = v_{ij}^{(0 \Omega)} / \check \tau
({\bf k}, {\bf \Omega}) \; ,
\label{L47}
\end{equation}
where $\check \tau ({\bf k}, {\bf \Omega}))$ is a relaxation time
analogous to $\tau (k)$ introduced with (\ref{Q7}). Now we assume
that the characteristic time of variations of $v_{ij}^{(0 \Omega)}
$ is much larger than $\check \tau$ so that we can drop the time
derivative $ \partial v_{ij}^{(0 \Omega)} / \partial t $ in
(\ref{L46}). Thinking of slow rotation, that is $ \Omega \check
\tau \ll 1 $, we expand $v_{ij}^{(0 \Omega)}$ in the form
\begin{equation}
v_{ij}^{(0 \Omega)} =  v_{ij}^{(01)} + v_{ij}^{(02)} \;,
\label{L48}
\end{equation}
where $v_{ij}^{(0 1)}$ and $v_{ij}^{(0 2)}$ are of the first and
second order in ${\bf \Omega}$, respectively. We further assume
that $\check \tau$ does not deviate markedly from $\tau$, that it
does not depend on the sign of $\Omega$ and therefore its
expansion with respect to ${\bf \Omega}$ possesses no linear term.
In that sense we put $\check \tau = \tau + O(\Omega^2)$, where the
last term is without interest for the following. In this way we we
obtain
\begin{equation}
v_{ij}^{(0, \, \mu)} = - \tilde \tau_{\ast} (\tilde M_{ijklm} \nabla_{m}
+ \tilde N_{ijkl}) v_{kl}^{(0, \mu -1)} \; ,
\quad {\tilde \tau_{\ast}}^{-1} = \tau^{-1} + 2 \nu k^2 \; ,
\quad \mu = 1,2 \,.
\label{L50}
\end{equation}
Identifying now $v_{kl}^{(00)}$ with $v_{kl}^{(0)}$ defined by (\ref{L18})
with $\mu^{(v)} = 0$ and neglecting again terms of higher than first order
in ${\bf \nabla}$ we find
\begin{equation}
v_{ij}^{(01)} = \frac{i \tilde \tau_{\ast}}{4 \pi k^{4}}
\varepsilon_{ijm} k_m \biggl( \frac{({\bf k} \cdot {\bf \Omega})}{k^2}
({\bf k} \cdot {\bf \nabla})
- ({\bf \Omega} \cdot {\bf \nabla}) \biggr)  W^{(v)}(k,{\bf R}) \; ,
\quad v_{ij}^{(02)} = 0 \; .
\label{L51}
\end{equation}
With (\ref{L51}) we arrive at
\begin{eqnarray}
v^{(0)}_{ij}({\bf k},{\bf R}) &=& \frac{1}{8 \pi k^2}
\biggl( P_{ij}({\bf k}) + \frac{i}{2 k^2} (k_i \nabla_j - k_j \nabla_i)
\nonumber \\
& & + \frac{2 i \tilde \tau_{\ast}}{k^2} \varepsilon_{ijm} k_m
\bigl( \frac{({\bf k} \cdot {\bf \Omega})}{k^2}({\bf k} \cdot {\bf
\nabla}) - ({\bf \Omega} \cdot {\bf \nabla}) \bigr) \biggr)
W^{(v)}(k,{\bf R}) \;. \label{L52}
\end{eqnarray}
Interestingly enough this coincides with (\ref{L18}) if we replace
there $\mu^{(v)}$ by $\tilde \mu^{(v)}$, defined by
\begin{equation}
\tilde \mu^{(v)} = - 2 \tilde \tau_{\ast}
\biggl( \frac{({\bf k} \cdot {\bf \Omega})}{k^2} ({\bf k} \cdot {\bf
\nabla})
- ({\bf \Omega} \cdot {\bf \nabla}) \biggr)  W^{(v)}(k,{\bf R}) \; ,
\label{L53}
\end{equation}
and cancel the terms with ${\bf \nabla} \tilde \mu^{(v)}$, which
are of second order in ${\bf \nabla}$. Note that the absence of
terms of higher than second order in ${\bf \Omega}$ in the
correlation tensor (\ref{L52}) is a consequence of the fact that
we restricted ourselves to small ${\bf \Omega} \check \tau$ and
used in that sense (\ref{L48}). The absence of terms of the second
order results from the assumption on the dependence of $\check
\tau$ on ${\bf \Omega}$.

As for the magnetic fluctuations of the original turbulence we
assume that they are independent of the rotation of the fluid and
show no helicity so that
\begin{equation}
m^{(0)}_{ij}({\bf k},{\bf R}) = \frac{1}{8 \pi k^2}
\biggl( P_{ij}({\bf k}) + \frac{i}{2 k^2} (k_{j} \nabla_{i} - k_{i}
\nabla_{j}) \biggr)
W^{(m)}(k,{\bf R}) \; .
\label{L55}
\end{equation}

We calculate now the $v_{ijk \cdots p}^{(\mu)}$, $m_{ijk \cdots
p}^{(\mu)}$, $\cdots$ defined by (\ref{B31})--(\ref{B32}) with
$v_{ij}$ replaced by $v_{ij}^{(0)}$ according to (\ref{L52}) and
find
\begin{eqnarray}
v_{ij}^{(\mu)} &=& \frac{1}{3} \delta_{ij} I^{(v \mu)} + \cdots \; ,
\label{Y21} \\
v_{ijk}^{(\mu)} &=& \frac{1}{12} ( \delta_{ik} \nabla_j - \delta_{jk}
\nabla_i ) I^{(v \mu)}
\nonumber \\
& & - \frac{1}{3} \varepsilon_{ijp} ( \delta_{pk} ({\bf \Omega}
\cdot {\bf \nabla} {\tilde I}^{(v \mu)}) - \frac{3}{5}
E_{pklm}^{(4)} \Omega_l \nabla_m {\tilde I}^{(v \mu)} ) \; ,
\label{Y22} \\
v_{ijkl}^{(\mu)} &=& \frac{1}{6} (\delta_{ij} \delta_{kl}
- \frac{3}{5} E_{ijkl}^{(4)} ) I^{(v \mu)} + \cdots \; ,
\label{Y23} \\
v_{ijklm}^{(\mu)} &=& \frac{1}{20}
( E_{iklm}^{(4)} \nabla_j - E_{jklm}^{(4)} \nabla_i ) I^{(v \mu)}
\nonumber \\
& & - \frac{1}{5} \varepsilon_{ijp} ( E_{pklm}^{(4)} ({\bf \Omega}
\cdot {\bf \nabla} {\tilde I}^{(v \mu)}) - \frac{5}{7}
E_{pklmqr}^{(6)} \Omega_q \nabla_r {\tilde I}^{(v \mu)} ) \; ,
\label{Y24} \\
v_{ijklmp}^{(\mu)} &=& \frac{1}{10} ( \delta_{ij} E_{klmp}^{(4)}
- \frac{5}{7} E_{ijklmp}^{(6)} ) I^{(v \mu)} + \cdots \; ,
\label{Y25}
\end{eqnarray}
In accordance with (\ref{Y14}) we have used here the definitions
\begin{equation}
I^{(v \mu)} = \int_0^{\infty} W^{(v)}(k,{\bf R}) \tau_{\ast}^{\mu}(k) \,d k
\;,
\quad {\tilde I}^{(v \mu)} = \int_0^{\infty} W^{(v)}(k,{\bf R})
\tau_{\ast}^{\mu}(k)
{\tilde \tau}_{\ast}(k) \,d k \; .
\label{Y26}
\end{equation}
The expression for $v_{ijklmp}^{(\mu)}$ turns into $v_{ijklmp}^{\prime}$
if $I^{(v \mu)}$ is replaced by $I^{(v) \prime}$ where
\begin{equation}
I^{(v) \prime} = \int_0^{\infty} W^{(v)}(k,{\bf R}) (d \tau_{\ast}^2 / d k)
k \,d k \; .
\label{Y27}
\end{equation}
Again terms indicated by $\cdots$ correspond to contributions of
second order in ${\bf \nabla}$ to ${\bf \cal{E}}$. Relations
analogous to (\ref{Y21})--(\ref{Y27}) apply for $m_{ij}^{(\mu)}$,
$m_{ijk}^{(\mu)}, \cdots $, too, but with ${\tilde I}^{(m \mu)} =
0$.

Calculating now the $a_{ij}^{(v)}$, $a_{ij}^{(m)}, \cdots$
according to (\ref{B26})--(\ref{B30}) and finally the
$\alpha_{ij}^{(v)}$, $\alpha_{ij}^{(m)}, \cdots$ according to
(\ref{La})--(\ref{Lb}) we arrive at
\begin{eqnarray}
\alpha_{ij}^{(v)}
&=&  \frac{2}{15} \left( 4 \delta_{ij} ({\bf \Omega} \cdot \bec{\nabla})
- \Omega_{i} \nabla_{j} - \Omega_{j} \nabla_{i}) \right) {\tilde I}^{(v1)}
\nonumber \\
& & + \frac{4}{15} \left( \delta_{ij} ({\bf \Omega} \cdot
\bec{\nabla}) - \frac{7}{8} (\Omega_{i} \nabla_{j} + \Omega_{j}
\nabla_{i}) \right) I^{(v2)}
\nonumber \\
& & + \frac{2}{15} ( \varepsilon_{ilm} \Omega_j +
\varepsilon_{jlm} \Omega_i ) \Omega_l \nabla_m I^{(v3)} \;,
\label{B47} \\
\alpha_{ij}^{(m)}
&=&  - \frac{1}{10} \left( \frac{8}{3} \delta_{ij} ({\bf \Omega} \cdot
\bec{\nabla})
+ \Omega_{i} \nabla_{j} + \Omega_{j} \nabla_{i} \right) I^{(m2)}
\nonumber \\
& & - \frac{2}{15} ( \varepsilon_{ilm} \Omega_j +
\varepsilon_{jlm} \Omega_i ) \Omega_l \nabla_m I^{(m3)} \;,
\label{B48} \\
\beta_{ij}^{(v)}
&=& \frac{1}{3} \delta_{ij} I^{(v1)}
- \frac{2}{5} (\delta_{ij} \Omega^2 + \frac{1}{3} \Omega_{i} \Omega_{j})
I^{(v3)} \;,
\label{B49} \\
\beta_{ij}^{(m)}
&=& \frac{2}{15} ( \delta_{ij} \Omega^2 - 3 \Omega_{i} \Omega_{j} ) I^{(m3)}
\;,
\label{B49a} \\
\bec{\gamma}^{(v)}
&=& \frac{1}{6} \bec{\nabla} ( I^{(v1)} - \frac{8}{5} \Omega^2 I^{(v3)} )
+ \frac{1}{6} {\bf \Omega} \times \bec{\nabla} I^{(v2)}
+ \frac{2}{15} {\bf \Omega} ( {\bf \Omega} \cdot \bec{\nabla} I^{(v3)} ) \;,
\label{B49b} \\
\bec{\gamma}^{(m)}
&=& - \frac{1}{6} \bec{\nabla} ( I^{(m1)} - \frac{8}{5} \Omega^2 I^{(m3)} )
+ \frac{1}{6} {\bf \Omega} \times \bec{\nabla} I^{(m2)}
- \frac{2}{15} {\bf \Omega} ( {\bf \Omega} \cdot \bec{\nabla} I^{(m3)} ) \;,
\label{B49c} \\
\bec{\delta}^{(v)}
&=& - \frac{1}{6} {\bf \Omega} I^{(v2)} \;,
\label{B49d} \\
\bec{\delta}^{(m)}
&=& \frac{1}{6} {\bf \Omega}  I^{(m2)} \;,
\label{B49e} \\
\kappa_{ijk}^{(v)}
&=&  - \frac{1}{6} ( \delta_{ij} \Omega_{k} + \delta_{ik} \Omega_{j})
( I^{(v2)} + \frac{2}{5} I^{(v) \prime} )
- \frac{2}{15} ( \varepsilon_{ijl} \Omega_l \Omega_k + \varepsilon_{ikl}
\Omega_l \Omega_j)
I^{(v3)} \; ,
\label{B50} \\
\kappa_{ijk}^{(m)}
&=&  - \frac{7}{30} ( \delta_{ij} \Omega_{k} + \delta_{ik} \Omega_{j})
( I^{(m2)} + \frac{2}{7} I^{(m) \prime} )
+ \frac{2}{15} ( \varepsilon_{ijl} \Omega_l \Omega_k + \varepsilon_{ikl}
\Omega_l \Omega_j)
I^{(m3)} \; .
\label{B51}
\end{eqnarray}

\subsection*{8.3. Specification to Kolmogorov Type Turbulence Spectra}

Let us now specify the original turbulence to be of Kolmogorov
type, i.e., to possess a  constant energy flux through the
spectrum, and consider the inertial range of wave numbers, $ k_{0}
\leq k \leq k_{d} ,$ where $ k_{0}^{-1} = l_{0} $ defines the
largest length scale and $ k_{d}^{-1} $ the dissipative scale of
the turbulence. In this range we have $ W^{(v)} = (q - 1) (\langle
{\bf u}^{(0)2} \rangle / k_{0}) (k / k_{0})^{-q} ,$ and $ W^{(m)}
$ analogously. Furthermore we put $ \mu^{(v)} = (q - 1) (\langle
{\bf u}^{(0)} \cdot (\bec{\nabla} \times {\bf u}^{(0)}) \rangle /
k_{0})  (k / k_{0})^{-q} $ but, by reasons connected with the
conservation of the magnetic helicity in the high--conductivity
limit, $ \mu^{(m)} = \langle {\bf b}^{(0)} \cdot (\bec{\nabla}
\times {\bf b}^{(0)}) \rangle \delta(k - k_0) $ . Finally we
assume that $\tau_{\ast} = {\tilde \tau}_{\ast} = 2 \tau_{0} (k /
k_{0})^{1-q}$, with $2\tau_{0}$ being a correlation (or turnover)
time for $k = k_0$. In all cases $q$ is a constant constrained by
$1 < q < 3$. Assuming that it is sufficient to take the integrals
in (\ref{Y14}) and (\ref{Y26})--(\ref{Y27}) over the inertial
range only and that $k_{0} / k_{d} \ll 1$ we obtain
\begin{eqnarray}
I^{(v \mu)} &=& \frac{2^{\mu}}{\mu +1} \langle {\bf u}^{(0)2} \rangle
\tau_{0}^{\mu} \;,
\quad J^{(v1)} = \langle {\bf u}^{(0)} \cdot (\bec{\nabla} \times {\bf
u}^{(0)}) \rangle
\tau_{0} \;,
\label{B65} \\
{\tilde I}^{(v1)} &=& I^{(v2)} \;,
\quad I^{(v) \prime} = - 2 (q - 1) I^{(v2)} \;,
\label{B65a} \\
I^{(m \mu)} &=& \frac{2^{\mu}}{\mu + 1} \frac{\langle {\bf b}^{(0)2}
\rangle}{\mu \rho}
\tau_{0}^{\mu} \;,
\quad J^{(m1)} =
\frac{\langle {\bf b}^{(0)} \cdot (\bec{\nabla} \times {\bf b}^{(0)})
\rangle}{\mu \rho}
\tau_{0} \;,
\label{B66} \\
I^{(m) \prime} &=& - 2 (q - 1) I^{(m2)} \;,
\label{B66a}
\end{eqnarray}
Note that the $I^{(v \mu)}$, $J^{(v1)}$, ${\tilde I}^{(v1)}$,
$I^{(m \mu)}$ and $J^{(m1)}$ are independent of $q$.

By the way, with the above specification of $\mu^{(m)}$ we have
simply $\langle {\bf b}^{(0)} \cdot (\bec{\nabla} \times {\bf
b}^{(0)}) \rangle = \langle {\bf a}^{(0)} \cdot {\bf b}^{(0)}
\rangle / l_{0}^{2}$. Here $\langle {\bf a}^{(0)} \cdot {\bf
b}^{(0)} \rangle$ is the magnetic helicity, where ${\bf a}^{(0)}$
is the vector potential of ${\bf b}^{(0)}$, \mbox{i. e.}, ${\bf
b}^{(0)} = \bec{\nabla} \times {\bf a}^{(0)}$. The factor
$\delta(k - k_0)$ in the function $\mu^{(b)}$ is chosen in order
to meet the realisability condition for the magnetic helicity
(see, e.g., Moffatt, 1978; Zeldovich et al. 1983).

\subsection*{8.4. Specific Results for Nonrotating Turbulence}

Let us now summarize and discuss our results. When speaking in
this context simply of contributions of velocity and magnetic
field fluctuations we refer always to the original turbulence and
so to ${\bf u}^{(0)}$ and ${\bf b}^{(0)}$. Starting with the case
of nonrotating turbulence we specify now (\ref{B44})--(\ref{B46})
by (\ref{B65})--(\ref{B66a}) and obtain
\begin{eqnarray}
\alpha_{ij} &=& \alpha \delta_{ij} \; ,
\quad \alpha = \frac{1}{3}
\biggl( \langle {\bf u}^{(0)} \cdot (\bec{\nabla} \times {\bf u}^{(0)})
\rangle
- \frac{\langle {\bf b}^{(0)} \cdot (\bec{\nabla} \times {\bf b}^{(0)})
\rangle}{\mu \rho}
\biggr) \tau_0 \; ,
\label{B69} \\
\beta_{ij} &=& \beta \delta_{ij} \; ,
\quad \beta = \frac{1}{3} \langle {\bf u}^{(0)2} \rangle \tau_{0} \; ,
\label{B70} \\
{\bec \gamma} &=& \frac{1}{6} \bec{\nabla} \biggl( \langle {\bf u}^{(0)2}
\rangle
- \frac{\langle {\bf b}^{(0)2} \rangle}{\mu\rho} \biggr) \tau_{0} \; ,
\label{B71} \\
{\bec \delta} &=& {\bf 0} \; ,
\quad {\bec \kappa} = {\bf 0} \; .
\label{B72}
\end{eqnarray}

We have an isotropic $\alpha$--effect, and $\alpha$ is a sum
$\alpha^{(v)} + \alpha^{(m)}$ of two contributions determined by
the kinematic helicity and the current helicity of the original
turbulence. Whereas the signs of $\alpha^{(v)}$ and the kinematic
helicity coincide, those of $\alpha^{(m)}$ and the current
helicity are opposite to each other. This is in agreement with
results, e.g., by Pouquet et al. 1976, by Zeldovich et al. 1983 or
by Vainshtein and Kichatinov 1983. The contributions of the
kinematic and the current helicities to the $ \alpha $-effect,
acting in opposite directions, may even compensate each other.

We also have an isotropic mean-field diffusivity $\beta$, and this
is determined only by the intensity of the velocity fluctuations
of the original turbulence. There is no contribution of the
magnetic fluctuations of the original turbulence. This again
agrees with results by Vainshtein and Kichatinov 1983.

An inhomogeneity of the original turbulence leads to a transport
of mean magnetic flux which corresponds to that by a mean velocity
$-{\bec \gamma}$. The expulsion of flux from regions with high
intensity of the velocity fluctuations has been sometimes
discussed as ``turbulent diamagnetism''; see, e.g., Zeldovich
1956, R\"{a}dler 1970 or Krause and R\"{a}dler 1980. Our result
shows, again in agreement with Vainshtein and Kichatinov 1983,
that magnetic fluctuations act in the opposite sense, that is, in
the sense of a ``turbulent paramagnetism''. The magnetic flux is
expelled from regions where $\langle {\bf u}^{(0)2} \rangle - (1/
\mu \rho) \langle {\bf b}^{(0)2} \rangle $ is higher, and pushed
into regions where it is lower compared to the surroundings. With
equipartition of kinetic and magnetic energy, $\langle {\bf
u}^{(0)2} \rangle = (1/ \mu \rho) \langle {\bf b}^{(0)2} \rangle$,
this effect vanishes.

\subsection*{8.5. Specific Results for Rotating Turbulence}

In the case of the rotating turbulence in which helicity occurs
only due to inhomogeneity and rotation we have
\begin{eqnarray}
\alpha_{ij} &=& \frac{16}{15} \biggl( \delta_{ij} {\bf \Omega^{\star} \cdot}
\bec{\nabla}
( \langle {\bf u}^{(0)2} \rangle
- \frac{1}{3} \frac{\langle {\bf b}^{(0)2} \rangle}{\mu \rho})
\nonumber \\
& & - \frac{11}{24}(\Omega^{\star}_i \nabla_j + \Omega^{\star}_j
\nabla_i) ( \langle {\bf u}^{(0)2} \rangle + \frac{3}{11}
\frac{\langle {\bf b}^{(0)2} \rangle}{\mu\rho} )
\nonumber \\
& & + \frac{1}{3} ( \varepsilon_{ilm} \Omega^{\star}_l
\Omega^{\star}_j + \varepsilon_{jlm} \Omega^{\star}_l
\Omega^{\star}_i ) \nabla_m ( \langle {\bf u}^{(0)2} \rangle -
\frac{\langle {\bf b}^{(0)2} \rangle}{\mu \rho} ) \biggr) \tau_{0}
\; .
\label{B80} \\
\beta_{ij} &=& \frac{1}{3} \biggl( \delta_{ij} ( \langle {\bf
u}^{(0)2} \rangle ( 1 - \frac{12}{5} {\Omega^{\star}}^2 ) +
\frac{4}{5} {\Omega^{\star}}^2 \frac{\langle {\bf b}^{(0)2}
\rangle}{\mu \rho} )
\nonumber \\
& & - \frac{4}{5} \Omega_i^{\star} \Omega_j^{\star} ( \langle {\bf
u}^{(0)2} \rangle + 3 \frac{\langle {\bf b}^{(0)2}
\rangle}{\mu\rho} ) \biggr) \tau_{0}\;.
\label{B81} \\
{\bec \gamma} &=& \frac{1}{6} \biggl( \bec{\nabla}( \langle {\bf
u}^{(0)2} \rangle - \frac{\langle {\bf b}^{(0)2}
\rangle}{\mu\rho}) (1 -  \frac{16}{5} {\Omega^{\star}}^2) +
\frac{4}{3} {\bf \Omega^{\star}} \times \bec{\nabla} ( \langle
{\bf u}^{(0)2} \rangle + \frac{\langle {\bf b}^{(0)2}
\rangle}{\mu\rho} )
\nonumber \\
& & + \frac{8}{5} {\bf \Omega^{\star}} ({\bf \Omega^{\star}} \cdot
\bec{\nabla} (\langle {\bf u}^{(0)2} \rangle - \frac{\langle {\bf
b}^{(0)2} \rangle}{\mu\rho})) \biggr) \tau_{0} \;.
\label{B82} \\
{\bec \delta} &=& - \frac{2}{9} \bigl( \langle {\bf u}^{(0)2} \rangle
- \frac{\langle {\bf b}^{(0)2} \rangle}{\mu\rho} \bigr)
{\bf \Omega}^{\star} \tau_{0} \;,
\label{B83} \\
\kappa_{ijk} &=& - \frac{2}{9} \biggl( (\delta_{ij}
\Omega_k^{\star} + \delta_{ik} \Omega_j^{\star}) (\langle {\bf
u}^{(0)2} \rangle + \frac{7}{5} \frac{\langle {\bf b}^{(0)2}
\rangle}{\mu\rho} - \frac{4 (q - 1)}{5} ( \langle {\bf u}^{(0)2}
\rangle + \frac{\langle {\bf b}^{(0)2} \rangle}{\mu\rho} ) )
\nonumber \\
& & + \frac{6}{5} (\varepsilon_{ijl} \Omega_l^{\star}
\Omega_k^{\star} + \varepsilon_{ikl} \Omega_l^{\star}
\Omega_j^{\star}) (\langle {\bf u}^{(0)2} \rangle - \frac{\langle
{\bf b}^{(0)2} \rangle}{\mu\rho}) \biggr) \tau_{0} \;. \label{B84}
\end{eqnarray}
Here ${\bf \Omega^{\star}}$ stands for ${\bf \Omega }\tau_{0}$ .
Our approximations are justified for $ |\Omega^{\star}| \ll 1$ only.
We recall that we have ignored all terms containing factors
$(\eta - \nu) k^2 \tau_{\ast}$ in (\ref{B24}) and (\ref{B25}).
The only influence of these terms on the above result would consist
in the occurrence of contributions to $\alpha_{ij}$ which, in comparison
to others, are smaller by a factor of the order
of the small quantity $(\eta - \nu) k_0^2 \tau_0$.
As far as contributions of the velocity fluctuations are concerned
the result (\ref{B80})--(\ref{B84}) is in qualitative agreement
with earlier results (e.g., R\"{a}dler 1980).

The $\alpha$--effect occurs now as a consequence of the
simultaneous presence of a rotation of the fluid and gradients in
the intensities of the velocity or magnetic fluctuations, and it
is clearly anisotropic. A rough measure of the $\alpha$--effect is
the trace of $\alpha_{ij}$. For this quantity we have
\begin{equation}
\alpha_{ii} = \frac{20}{9} ({\bf \Omega^{\star} \cdot}
\bec{\nabla}) \biggl( \langle {\bf u}^{(0)2} \rangle
- \frac{3}{5} \frac{\langle {\bf b}^{(0)2} \rangle}{\mu \rho} \biggr)
\tau_{0}  \; .
\label{B85}
\end{equation}
At least with this quantity the effect of the velocity
fluctuations is again diminished by the magnetic fluctuations.

Like the $\alpha$--effect the mean--field diffusivity, too, is
in general anisotropic. Interestingly enough, in contrast to the case
of nonrotating turbulence $\beta_{ij}$ is now no longer independent
of the magnetic fluctuations. For equipartition of kinetic and
magnetic energy, $ \langle {\bf u}^{(0)2} \rangle = (1/ \mu \rho)
\langle {\bf b}^{(0)2} \rangle ,$ the mean-field diffusivity is
again isotropic.
Even in the general case the tensor $\beta_{ij}$ has no other
non--zero elements than diagonal ones.
If ${\bec{\Omega}}^{\star}$ is parallel to the $x_3$--axis
we have
\begin{eqnarray}
\beta_{11} = \beta_{22} &=& \frac{1}{3} \langle {\bec{u}}^{(0)2} \rangle
    \big(1 - \frac{12}{5} {\bec{\Omega}^{\star}}^2 \big)
    + \frac{4}{15} \frac{\langle {\bec{b}}^{(0)2} \rangle}{\mu \rho}
    {\bec{\Omega}^{\star}}^2
\nonumber\\
\beta_{33} &=& \frac{1}{3} \langle {\bec{u}}^{(0)2} \rangle
    \big(1 - \frac{16}{5} {\bec{\Omega}^{\star}}^2 \big)
    - \frac{8}{15} \frac{\langle {\bec{b}}^{(0)2} \rangle}{\mu \rho}
    {\bec{\Omega}^{\star}}^2 \, .
\label{B85a}
\end{eqnarray}
In the absence of magnetic fluctuations not caused by the mean magnetic
field,
i.e. $\bec{b}^{(0)} = {\bf 0}$, all elements of $\beta_{ij}$ decrease
with growing $|{\bec{\Omega}}^{\star}|$.

Compared to the case of nonrotating turbulence, the vector ${\bec
\gamma}$ describing the transport of mean magnetic flux has an
additional term perpendicular to the rotation axis. This term
contains no longer the gradient of $\langle {\bf u}^{(0)2} \rangle
- (1/ \mu \rho) \langle {\bf b}^{(0)2} \rangle$ but that of
$\langle {\bf u}^{(0)2} \rangle + (1/ \mu \rho) \langle {\bf
b}^{(0)2} \rangle$, that is, it does not vanish with $\langle {\bf
u}^{(0)2} \rangle = (1/ \mu \rho) \langle {\bf b}^{(0)2} \rangle$.

In contrast to the case of nonrotating turbulence even in the
absence of gradients of the turbulence intensities, the vector
${\bec \delta}$ is no longer equal to zero. So we rediscover the
contribution to the mean electromotive force proportional to
$\bec{\Omega} \times ( {\bf \nabla} \times \overline{\bf B})$ ,
which has been sometimes discussed
as ``$\bec{\Omega} \times {\bf J}$--effect''.
We note that $\bec{\Omega} \times ( {\bf \nabla} \times \overline{\bf B})
= - (\bec{\Omega} \cdot {\bf \nabla}) \overline{\bf B}
+ {\bf \nabla} (\bec{\Omega} \cdot \overline{\bf B})$
and that the term ${\bf \nabla} (\bec{\Omega} \cdot \overline{\bf B})$
plays no part in the induction equation for $\overline{\bf B}$
as long as the coefficient connecting $\bec{\delta}$ and $\bec{\Omega}$
does not depend on space--coordinates.
We recall that an electromotive force proportional to
$\bec{\Omega} \times ( {\bf \nabla} \times \overline{\bf B})$ is,
even in the absence of an $ \alpha $--effect, in combination
with a differential rotation capable of dynamo action; see, e.g.,
R\"{a}dler 1969, 1970, 1980, 1986, Roberts 1972 and Moffatt and
Proctor 1982. With respect to ${\bec \delta}$ velocity and
magnetic fluctuations act again in the opposite sense, and ${\bec
\delta}$ vanishes with $\langle {\bf u}^{(0)2} \rangle = (1/ \mu
\rho) \langle {\bf b}^{(0)2} \rangle$. For a rotating turbulence,
again even in the absence of gradients of the turbulence
intensities, also $\kappa_{ijk}$ is unequal to zero.

We recall that we determined the correlation tensor $v^{(0)}_{ij}$
used for the calculation of (\ref{B80}) -- (\ref{B84}) under the
assumption that the expansion of the relaxation time $\check \tau$
with respect to ${\bf \Omega}$ has no linear term. As a
consequence $v^{(0)}_{ij}$ contains no terms of second order in
${\bf \Omega}$. It can be followed up easily, however, that a
deviation from this assumption would change in (\ref{B80}) --
(\ref{B84}) nothing else than numerical factors of the terms of
second order in ${\bf \Omega^{\star}}$, and it would leave
(\ref{B85}) unchanged.

Our results (\ref{B80}) -- (\ref{B82}) are different from those
given in papers by Kichatinov (1991), R\"{u}diger \& Kichatinov
(1993) and Kichatinov et al. (1994). In the derivations of these
authors more restrictive assumptions have been used. In
particular, they dropped the nonlinear terms in the equations for
the velocity and magnetic field fluctuations from the very
beginning, which is only valid either for small hydrodynamic and
magnetic Reynolds numbers or for high magnetic Reynolds and small
Strouhal numbers. They further assumed that $\nu = \eta =
l_{c}^{2} / \tau_{c}$, where $\nu$ and $\eta$ are kinematic
viscosity and magnetic diffusivity, and $l_{c}$ and $\tau_{c}$
correlation length and time of the turbulent velocity field. This
in turn applies only if the hydrodynamic and magnetic Reynolds
numbers are of the order of unity. Our results, which are derived
on another way, apply for large hydrodynamic and magnetic Reynolds
numbers.

\subsection*{8.6. Implications for Mean--Field Dynamo Models}

Let us add some remarks on the possibilities of dynamo action
of the induction effects described by (\ref{B80}) -- (\ref{B84}).
For the sake of simplicity we consider an axisymmetric dynamo model.
We use corresponding cylindrical co--ordinates $r$, $\varphi$ and $z$.
The mean motion is assumed to consist in a differential rotation,
\begin{equation}
\overline{\bf U} = \omega {\bf e}_z \times {\bf r} \, ,
\label{N01}
\end{equation}
where the angular velocity $\omega$ may depend on $r$ and $z$,
${\bf e}_z$ is the unit vector in $z$--direction
and ${\bf r}$ the radius vector.
Further, again for simplicity, only the contributions to $\bec{\cal{E}}$
which are linear in ${\bf \Omega^{\star}}$ and independent of ${\bf
b}^{(0)}$
are taken into account.
We write
\begin{eqnarray}
\bec{\cal{E}} &=& - \alpha_0 (\bec{g} \cdot \bec{\Omega}^{\star})
\overline{\bf B}
   + \alpha_1 \big( (\bec{\Omega}^{\star} \cdot \overline{\bf B}) \, \bec{g}
   + (\bec{g} \cdot \overline{\bf B}) \, \bec{\Omega}^{\star} \big)
   - \beta_0 {\bf \nabla} \times \overline{\bf B}
\nonumber\\
   && - \gamma_0 \, \bec{g} \times \overline{\bf B}
   - \gamma_1 \big( (\bec{\Omega}^{\star} \cdot \overline{\bf B}) \, \bec{g}
   - (\bec{g} \cdot \overline{\bf B}) \, \bec{\Omega}^{\star} \big)
\label{N02}\\
   && - \delta_0 \big( (\bec{\Omega}^{\star} \cdot {\bf \nabla}) \,
\overline{\bf B}
   - {\bf \nabla} (\bec{\Omega}^{\star} \cdot \overline{\bf B}) \big)
   + \kappa_0 \big( (\bec{\Omega}^{\star} \cdot {\bf \nabla}) \,
\overline{\bf B}
   + {\bf \nabla} (\bec{\Omega}^{\star} \cdot \overline{\bf B}) \big) \,
\nonumber
\end{eqnarray}
where $\alpha_0 = \frac{24}{11} \alpha_1 = \frac{16}{15} \langle \bec{\bf
u}^{(0)2} \rangle \tau_0$,
$\beta_0 = \frac{1}{3} \langle \bec{\bf u}^{(0)2} \rangle \tau_0$,
$\gamma_0 = \frac{3}{4} \gamma_1 = \frac{1}{6} \langle \bec{\bf u}^{(0)2}
\rangle \tau_0$,
$\delta_0 = \frac{2}{9} \langle \bec{\bf u}^{(0)2} \rangle \tau_0$,
$\kappa_0 = \frac{2}{9} \big(1 - \frac{4(q - 1)}{5} \big) \langle \bec{\bf
u}^{(0)2} \rangle \tau_0$
and $\bec{g} = {\bf \nabla} \langle \bec{\bf u}^{(0)2} \rangle /
\langle \bec{\bf u}^{(0)2} \rangle$.
$\bec{\Omega}^{\star}$ is assumed to be parallel to the $z$--axis,
$\bec{\Omega}^{\star} = \Omega^{\star} {\bf e}_z$ with $\Omega^{\star} > 0$.

We split $\overline{\bf B}$ into its poloidal and toroidal part,
${\overline{\bf B}}^{P}$ and ${\overline{\bf B}}^T$,
and represent them in the form
\begin{equation}
{\overline{\bf B}}^P = {\bf \nabla} \times (A {\bf e}_\varphi) \, , \quad
    {\overline{\bf B}}^T = B {\bf e}_\varphi \, ,
\label{N03}
\end{equation}
where $A$ and $B$ are two functions of $r$, $z$ and $t$,
and ${\bf e}_\varphi$ is the unit vector in $\varphi$--direction.
We assume that the differential rotation is so strong that concerning
the generation of ${\overline{\bf B}}^T$
from ${\overline{\bf B}}^P$ all contributions to $\bec{\cal{E}}$
except the $\beta_0$--term can be neglected.
Considering $\eta$ again as constant and starting from the induction
equation (\ref{S4}) for $\overline{\bf B}$ we arrive at
\begin{eqnarray}
\eta_{\mathrm{m}} \Delta' A - \alpha_0 \Omega^{\star} g_z B
    + \frac{\gamma_0}{r} \bec{g} \cdot {\bf \nabla} (r A)
    - (\delta_0 - \kappa_0) \Omega^{\star} \frac{\partial B}{\partial z}
    - \frac{\partial A}{\partial t} &=& 0
\nonumber\\
\eta_{\mathrm{m}} \Delta' B - \big( \frac{\partial \omega}{\partial r}
\frac{\partial}{\partial z}
    - \frac{\partial \omega}{\partial z} \frac{\partial}{\partial r} \big)
(r A)
    - \frac{\partial B}{\partial t} &=& 0 \, ,
\label{N04}
\end{eqnarray}
where $\eta_{\mathrm{m}} = \eta + \beta_0$
and
$$\Delta' f = \frac{\partial}{\partial r} \big( \frac{1}{r}
\frac{\partial}{\partial r} (r f) \big) + \frac{\partial^2
f}{\partial z^2} \; .$$

Let us restrict ourselves to a local analysis of these equations,
that is, to an investigation in some finite region of the $rz$--plane only.
For this purpose we assume that there the coefficients $\alpha_0$,
$\gamma_0$,
$\delta_0$ and $\kappa_0$ as well as $\bec{g}$ are constant.
Further we assume that there $\omega$ depends linearily on $r$ and is
independent of $z$,
that is, the quantity $G = r \partial \omega / \partial r$ is also constant,
and $\partial \omega /\partial z = 0$.
We use the ansatz
\begin{equation}
(A, B) = \mbox{Re} \big\{ (A_0, B_0) J_1 (k_r r) \exp(\mbox{i} k_z
z + \lambda t) \big\} \, , \label{N05}
\end{equation}
where $A_0$ and $B_0$ are complex constants, $J_1$ is the first--order
Bessel function
of first kind,
$k_r$ and $k_z$ are real constants and $\lambda$ is a complex constant.
If $g_r = 0$ this ansatz reduces (\ref{N04}) to a system of two linear
homogeneous
algebraic equations for $A_0$ and $B_0$.
We are interested in non-trivial solutions only and have therefore to
require
that the determinant of this system vanishes.
This leads to
\begin{equation}
\lambda = - \eta_{\mathrm{m}} k^2 + \mbox{i} \frac{\gamma_0 g_z k_z}{2}
    \pm \sqrt{- G \Omega^{\star}( \mbox{i} \alpha_0 g_z k_z + (\delta_0 -
\kappa_0) k^2_z )
    - (\frac{\gamma_0 g_z k_z}{2})^2}
\label{N06}
\end{equation}
whith $k^2 = k^2_r + k^2_z$.
If $g_r \not= 0$ this reduction works only in the limit $k_r \to 0$,
and then (\ref{N06}) applies with $k^2 = k^2_z$.

Solutions of the equations (\ref{N04}) with a $\lambda$ possessing
a non--negative real part correspond to non--decaying mean
magnetic fields. We first consider the case $\bec{g} \not= {\bf
0}$, in which we have an $\alpha$--effect. As can be easily seen
from (\ref{N06}) values of $\lambda$ with non--negative real part
and non-vanishing imaginary part occur if $k$ is sufficiently
small. They correspond to undamped dynamo waves travelling parallel
to the $z$--axis.
Let us further proceed to the case $\bec{g} = {\bf 0}$,
in which there is no $\alpha$--effect.
If then $- G \Omega^{\star} (\delta_0 - \kappa_0) > 0$, the quantity
$\lambda$
is real, and it takes non--negative values for sufficiently small
$k$. That is, even in the absence of an $\alpha$--effect dynamo
action proves be possible due to combination of differential
rotation with the $\delta$ or $\kappa$--effect. This includes the
possibility of such dynamos with $\delta$--effect mentioned above.
We see now that $\delta$ and $\kappa$--effect are in competition
and that they compensate each other if $\delta_0 = \kappa_0$. We
note that under our assumptions $\delta_0 - \kappa_0 =
\frac{8}{45} (q - 1) \langle \bec{\bf u}^{(0)2} \rangle$.
Since $q > 1$ we may conclude that the dynamo requires $G < 0$,
that is $\partial \omega / \partial r < 0$.

Of course, the results of the local analysis of the dynamo equations
should be confirmed by solving them in all conducting space
using proper boundary conditions.
This has been done so far in the investigations referred to above for dynamo
models
involving differential rotation and $\delta$--effect.

\bigskip

\section*{9. CONCLUDING REMARKS}

In this paper we have shown a procedure to calculate the mean
electromotive force $\bec{\cal{E}}$ for a magnetohydrodynamic
turbulence. The bounds of its applicability result mainly from the
use of a closure assumption for the deviation of the turbulence
from that for zero magnetic field and zero rotation. As explained
above it can only be justified for sufficiently small mean
magnetic fields and small rotation rates of the fluid. For
simplicity we have restricted ourselves to the case in which there
is, apart from the rotation, no mean motion. Specific results have
been derived for the limit in which the mean electromotive force
is linear in the mean magnetic field and a Kolmogorov-type
turbulence.

Let us compare results obtained in the kinematic approach on the
basis of the second-order correlation approximation, or
first-order smoothing, for the high-conductivity limit with
results of our procedure. Take as a simple example the case of
homogeneous and isotropic turbulence. Then we have $ \bec{\cal E}
= - \alpha \overline{\bf B} - \beta \bec{\nabla} \times
\overline{\bf B} $ for sufficiently weak variations of $
\overline{\bf B} $ in space and in time. In the kinematic approach
under the conditions mentioned we have
\begin{eqnarray}
\alpha &=&  {1 \over 3} \int_{0}^{\infty} \langle {\bf u}({\bf x},t)
\cdot (\bec{\nabla} \times {\bf u}({\bf x},t-\tau)) \rangle \,d \tau \;,
\label{B90} \\
\beta  &=& {1 \over 3} \int_{0}^{\infty} \langle {\bf u}({\bf x},t)
\cdot {\bf u}({\bf x},t-\tau)) \rangle \,d \tau \; .
\label{B91}
\end{eqnarray}
This is often expressed in the form
\begin{eqnarray}
\alpha = {1 \over 3}  \langle {\bf u} \cdot (\bec{\nabla} \times
{\bf u}) \rangle \tau_{\rm corr}^{(\alpha)} \; ,
\quad
\beta = {1 \over 3}  \langle {\bf u}^{2} \rangle \tau_{\rm corr}^{(\beta)}
\;,
\label{B92}
\end{eqnarray}
with properly defined correlation times $ \tau_{\rm
corr}^{(\alpha)} $ and $ \tau_{\rm corr}^{(\beta)} .$ The validity
of these results can only be readily justified under the condition
$ u \tau / l \ll 1 ,$ where $ u ,$ $ l ,$ and $ \tau $ are typical
values of the velocity and of the length and time scales of the
velocity field. This condition, however, is problematic in view of
applications to realistic situations, and the validity of these
results beyond this condition is questionable. Basically it is
possible to improve the approximation by including higher-order
terms in $ {\bf u} $ but this is very tedious.

The results (\ref{B92}) have the structure of our result given by
(\ref{B69}), specified by $ {\bf b}^{(0)} = \bf{0} $, and our
result (\ref{B70}). The validity of our results, however, is not
restricted by a condition like $ u \tau / l \ll 1 ,$ but only by
the applicability of the $ \tau $-approximation (\ref{Q7}). That
is, we have in any case a much wider range of validity.

We point out that despite the formal similarity of the mentioned
results gained in the kinematic approach and those derived here,
there is a basic difference between them. In the first case we
have originally, that is in (\ref{B90}) and (\ref{B91}),
correlations between values of $\bf u$ taken at different times,
which are often as in (\ref{B92}) expressed by $\bf u$ at a given
time and a correlation time, but in the second case we consider
from the very beginning only correlations between values
of $\bf u$ at the same time.
In that sense there is no simple connection between the two kind of results.

We may use the framework explained above also beyond the limit of
very small mean magnetic fields and study, for example, $
\bec{\alpha} $ or $ \bec{\beta} $-quenching at least for not too
strong fields. Then, of course, $ v_{ij} $ and $ m_{ij} $ can no
longer be replaced by $ v^{(0)}_{ij} $ and $ m^{(0)}_{ij} $.
Instead we have to derive equations for $ v_{ij} $ and $ m_{ij} $
corresponding to equation (\ref{Q3}) for $ \chi_{ij} $, insert
their solutions depending on $ \overline{\bf B} $ in $ I_{ij} $ in
(\ref{B21}) and follow the above pattern of the determination of $
a_{ij} $ and $ b_{ijk} ,$ or $ \bec{\alpha} ,$ $ \bec{\beta} ,
\ldots $. By the way, then even in the case $ {\bf b}^{(0)} =
\bf{0} $ there are contributions of $ m_{ij} $ to these
coefficients which, of course, vanish like $ m^{(0)}_{ij} $ with $
\overline{\bf B} $.

In this context it suggests itself to study in addition to the
mean electromotive force $ \bec{\cal{E}} $ also the mean
ponderomotive force $ \bec{\cal{F}} $. It has a part independent
of $ \overline{\bf B} $, which can be calculated with the help of
$ v^{(0)}_{ij} $ and $ m^{(0)}_{ij} $ only. Its general form  can
be derived on the basis of solutions of the equations for $ v_{ij}
$ and $ m_{ij} $ mentioned.

\bigskip
\bigskip
{\bf Acknowledgment}

We thank Dr. M. Rheinhardt for inspiring discussions. We
acknowledge support from INTAS Program Foundation (Grant No.
99-348). N.K. and I.R. also thank the Astrophysical Institute of
Potsdam for its hospitality.

\appendix
\section{Relations with $\varepsilon_{ijk}$}

For the derivation of (\ref{B8}) it is useful to know the relation
\begin{equation}
\varepsilon_{ijk} \Omega_k
+ ( \varepsilon_{ikl} k_j - \varepsilon_{jkl} k_i ) \frac{k_k \Omega_l}{k^2}
= \varepsilon_{ijk} k_k \frac{({\bf k} \cdot {\bf \Omega})}{k^2} \; ,
\label{X1}
\end{equation}
which applies to arbitrary vectors ${\bf k}$ and ${\bf \Omega}$.

In view of the derivation of (\ref{B40}) we recall the identity
\begin{eqnarray}
\varepsilon_{ijk} \varepsilon_{lmn}
&=&  \delta_{il} \delta_{jm} \delta_{kn}
    + \delta_{in} \delta_{jl} \delta_{km}
    + \delta_{im} \delta_{jn} \delta_{kl}
\nonumber \\
  & & - \delta_{in} \delta_{jm} \delta_{kl}
    - \delta_{il} \delta_{jn} \delta_{km}
    - \delta_{im} \delta_{jl} \delta_{kn} \; .
\label{X2}
\end{eqnarray}

\section{Derivation of equation (\ref{Q3})}

In the calculations of $ \partial \chi_{ij} / \partial t $ on the
basis of the equations (\ref{Q1}) and (\ref{Q2}) contributions to
this quantity occur which have, e.g., the form of
\begin{eqnarray}
X_{ij}({\bf k},{\bf R}) = \int \langle \hat S_{i}({\bf u},
\overline{{\bf B}}; {\bf k} + {\bf K}/2) \hat u_j( -{\bf k} + {\bf
K}  / 2 ) \rangle \exp{(i {\bf K \cdot R}) } \,d^{3} K
\nonumber \\
= i \int (k_{k} + K_{k} / 2) \langle \hat u_i ({\bf k} + {\bf  K}/
2 - {\bf Q}) \hat u_j(- {\bf k} + {\bf  K}/ 2) \rangle
\hat{\overline{B}}_{k}({\bf Q}) \exp{(i {\bf K \cdot R}) } \,d^{3}
K \,d^{3} Q \; . \label{B54}
\end{eqnarray}
In the last expression we may change the sequence of integration
so that the inner integral is over $ {\bf  K} $ and the outer
integral is over $ {\bf  Q} .$ In the inner integral we may
further change the integration variable $ {\bf K} $ into $ {\bf
K}- {\bf Q} ,$ denoted by $ {\bf  K}' $ in the following. In this
way, and using $ Q_k \hat{\overline{B}}_k = 0 ,$ we obtain
\begin{eqnarray}
X_{ij}({\bf k},{\bf R})
&=& i \int (k_{k} + K'_{k} / 2) \langle \hat u_i ({\bf k} - {\bf Q}/ 2 +
{\bf  K}'/ 2)
\hat u_j(- {\bf k} + {\bf Q}/ 2
\nonumber \\
& & + {\bf  K}'/ 2) \rangle \hat{\overline{B}}_k({\bf Q}) \exp[i(
{\bf K' \cdot R + Q \cdot R})] \,d^{3} K' \,d^{3} Q \; .
\label{B55}
\end{eqnarray}
Remembering the definition of $v_{ij}(\bf{k, R})$ we can rewrite this into
\begin{eqnarray}
X_{ij}({\bf k},{\bf R}) =  \int \left[i k_{k} v_{ij}({\bf k - Q} /
2,{\bf R}) + \frac{1}{2} \left( {\partial v_{ij}({\bf k - Q} /
2,{\bf R}) \over \partial R_k} \right) \right]
\hat{\overline{B}}_k({\bf Q}) \exp{(i {\bf Q \cdot R}) } \,d^{3} Q
\; . \label{B56}
\end{eqnarray}
The fact that $ \overline{\bf B} $ varies only on large scales,
that is, $ \hat{\overline{\bf B}} $ is only non-zero for certain
small $ \vert {\bf Q} \vert ,$ suggests to use the Taylor
expansion
\begin{eqnarray}
v_{ij}({\bf k} - {\bf Q}/ 2,{\bf R}) \simeq v_{ij}({\bf k},{\bf
R}) - \frac{1}{2} \left({\partial v_{ij}({\bf k},{\bf R}) \over
\partial k_k} \right) Q_k  + O({\bf Q}^2) \; .
\label{B57}
\end{eqnarray}
This yields
\begin{eqnarray}
X_{ij}({\bf k},{\bf R}) \simeq [i({\bf k }\cdot \overline{\bf B})
+ \frac{1}{2} (\overline{\bf B} \cdot \bec{\nabla}) ] v_{ij}({\bf k},{\bf
R})
- k_{k} v_{ijl} ({\bf k},{\bf R}) \overline{B}_{k,l} \;,
\label{B58}
\end{eqnarray}
where $ v_{ijl} = \frac{1}{2} \partial v_{ij} / \partial k_l .$
According to our assumption a term of the second-order in $
\bec{\nabla} $ was neglected. The contributions to the Taylor
expansion indicated by $ O({\bf Q}^2) $ only lead to terms of
higher order in $ \bec{\nabla} $ and need not to be considered.

\newpage
\leftline{\bf References}

Blackman, E. G. \& Brandenburg, A., ``Dynamic nonlinearity in
large scale dynamos with shear,'' {\it Astrophys. J.} {\bf 579},
Issue: November 1, (2002); (astro-ph/0204497).

Brandenburg, A., Jennings, R. L., Nordlund, A., Rieutord, M.,
Stein, R. F. \& Tuominen, I., ``Magnetic structures in a dynamo
simulation", {\it J. Fluid Mech.} {\bf 306}, 325-352 (1996).

Cattaneo, F. \& Hughes, D. W., ``Nonlinear saturation of the
turbulent $\alpha$ effect", {\it Phys. Rev. E} {\bf 54},
R4532-R4535 (1996).

Childress, S. \& Gilbert, A.,  {\it Stretch, Twist, Fold: The Fast
Dynamo}, Springer-Verlag, Berlin (1995).

Field, G. B., Blackman, E. G. \& Chou, H., ``Non-linear alpha effect
in dynamo theory",  {\it Astrophys. J.} {\bf 513}, 638-651 (1999).

Gruzinov, A.  \& Diamond, P. H., ``Self-consistent theory of
mean-field electrodynamics", {\it Phys. Rev. Lett.} {\bf 72},
1651-1653 (1994).

Kazantsev, A. P., ``Enhancement of a magnetic field by a
conducting fluid", {\it Sov. Phys. JETP}, Engl. Transl., {\bf 26},
1031-1039 (1968).

Kichatinov, L. L., ``On mean-field magnetohydrodynamics in an
inhomogeneous medium", {\it Magnitnaya Gidrodinamika} {\bf 3},
67-73 (1982)(in Russian).

Kichatinov, L. L., ``Turbulent transport of magnetic fields in a
highly conducting rotating fluid and the solar cycle", {\it
Astron. Astrophys.} {\bf 243}, 483-491 (1991).

Kichatinov, L. L., R\"{u}diger \& Pipin V. V., ``Turbulent
viscosity, magnetic diffusivity, and heat conductivity under the
influence of rotation and magnetic field", {\it Astron. Nachr.}
{\bf 315}, 157-170 (1994).

Kleeorin, N., Mond, M. \& Rogachevskii, I., ``Magnetohydrodynamic
turbulence in the solar convective zone as a source of
oscillations and sunspots formation", {\it Astron. Astrophys.}
{\bf 307}, 293-309 (1996).

Kleeorin N., Moss D., Rogachevskii I. \& Sokoloff D., ``The role of
magnetic helicity transport in nonlinear galactic dynamos", {\it
Astron. Astrophys.} {\bf 387}, 453-462 (2002a).

Kleeorin, N., Rogachevskii, I. \& Ruzmaikin, A., ``Magnetic force
reversal and instability in a plasma with advanced
magnetohydrodynamic turbulence", {\it Soviet Physics-JETP} {\bf
70}, 878-883 (1990).

Kleeorin N., Rogachevskii I. \& Sokoloff D., ``Magnetic
fluctuations with zero mean field in a random fluid flow with a
finite correlation time and a small magnetic diffusion", {\it
Phys. Rev. E} {\bf 65}, 036303, 1-7 (2002b).

Krause, F. \& R\"{a}dler, K.-H., ``Elektrodynamik der mittleren
Felder in turbulenten leitenden Medien und Dynamotheorie", in R.
Rompe and M. Steebeck (eds.), {\it Ergebnisse der Plasmaphysik und
Gaselektronik}, Akademie--Verlag, Berlin, 1-154 (1971).

Krause, F. \& R\"{a}dler, K.-H., {\it Mean-Field
Magnetohydrodynamics and  Dynamo Theory}, Pergamon Press, Oxford
(1980).

Kulsrud, R., ``A critical review of galactic dynamos", {\it Ann.
Rev. Astron. Astrophys.} {\bf 37}, 37-64 (1999).

Meneguzzi, M.,  Frisch, U. \& Pouquet, A., ``Helical and
nonhelical turbulent dynamos", {\it Phys. Rev. Lett.}, {\bf 41},
1060-1064 (1981).

McComb, W. D., {\it The Physics of Fluid Turbulence}, Clarendon
Press, Oxford (1990).

Moffatt, H. K., {\it Magnetic Field Generation in Electrically
Conducting Fluids}, Cambridge Univ. Press, New York (1978).

Moffatt, H. K. \& Proctor, M. R. E., ``The role of the helicity
spectrum function in turbulent dynamo theory", {\it Geophys.
Astrophys. Fluid Dyn.} {\bf 21}, 265-283 (1982).

Monin, A. S. \& Yaglom, A. M., {\it Statistical Fluid Mechanics},
{\bf 2}, MIT Press, Cambridge/Massachusetts (1975).

Nordlund, A., Brandenburg, A., Jennings, R. L., Rieutord, M.,
Ruokolainen, J., Stein, R. F. \& Tuominen, I., ``Dynamo action in
stratified convection with overshoot", {\it Astrophys. J.} {\bf
392}, 647-652 (1992).

Orszag, S. A.`, ``Analytical theories of turbulence", {\it J. Fluid
Mech.} {\bf 41}, 363-386 (1970).

Parker, E., {\it Cosmical Magnetic Fields}, Oxford Univ. Press,
New York (1979).

Pouquet, A., Frisch, U. \& Leorat, J., ``Strong MHD turbulence and
the nonlinear dynamo effect", {\it J. Fluid Mech.} {\bf 77},
321-354 (1976).

R\"{a}dler, K.--H., ``Zur Elektrodynamik turbulenter bewegter
leitender Medien II. Turbulenzbedingte Leitf\"{a}higkeits- und
Permeabilit\"{a}ts\"anderungen", {\it Z. Naturforsch.} {\bf 23a},
1851-1860 (1968).

R\"{a}dler, K.--H., ``\"Uber eine neue M\"oglichkeit eines
Dynamomechanismus in turbulenten leitenden Medien", {\it
Monatsber. Dtsch. Akad. Wiss. Berlin} {\bf 11}, 272-279 (1969).

R\"{a}dler, K.--H., ``Untersuchung eines Dynamomechanismus in
leitenden Medien", {\it Monatsber. Dtsch. Akad. Wiss. Berlin} {\bf
12}, 468-472 (1970).

R\"{a}dler, K.--H., ``Mean--field magnetohydrodynamics as a basis
of solar dynamo theory", in V. Bumba and J. Kleczek (eds.), {\it
Basic Mechanisms of Solar Activity}, D. Reidel Publ. Company,
Dordrecht, pp. 323-344 (1976).

R\"{a}dler, K.--H., ``Mean-field approach to spherical dynamo
models", {\it Astron. Nachr.} {\bf 301}, 101-129 (1980).

R\"{a}dler, K.--H., ``On the mean--field approach to spherical
dynamo models",  In A. M. Soward (ed.), {\it Stellar and Planetary
Magnetism}, Gordon and Breach Publ., New York (1983).

R\"{a}dler, K.--H., ``Investigations of spherical kinematic
mean--field dynamo models", {\it Astron. Nachr.} {\bf 307}, 89-113
(1986).

Roberts, P. H., ``Kinematic dynamo models", {\it Phil. Trans. R.
Soc. London Ser. A} {\bf 272}, 663-703 (1972).

Roberts, P. H. \& Soward, A. M., ``A unified approach to mean field
electrodynamics", {\it Astron. Nachr.} {\bf 296}, 49-64 (1975).

Rogachevskii, I. \&  Kleeorin, N., ``Intermittency and anomalous
scaling for magnetic fluctuations", {\it Phys. Rev. E} {\bf 56},
417-426 (1997).

Rogachevskii, I. \&  Kleeorin, N., ``Nonlinear turbulent magnetic
diffusion and mean-field dynamo", {\it Phys. Rev. E} {\bf 64},
056307, 1-14 (2001).

R\"{u}diger, G. \& Kichatinov, L. L., ``Alpha--effect and
alpha--quenching", {\it Astron. Astrophys.} {\bf 269}, 581-588
(1993).

Seehafer, N., ``Nature of the $\alpha$ effect in
magnetohydrodynamics", {\it Phys. Rev. E} {\bf 53}, 1283-1286
(1996).

Vainshtein, S. I. \& Kichatinov, L. L., ``The macroscopic
magnetohydrodynamics of inhomogeneously turbulent cosmic plasmas",
{\it Geophys. Astrophys. Fluid Dynamics} {\bf 24}, 273-298 (1983).

Zeldovich, Ya. B., ``The magnetic field in the two-dimensional
motion of a conducting turbulent fluid", {\it Zh. Eksp. Teor.
Fiz.} {\bf 31}, 154-156 (1956); {\it Sov. Phys. JETP} {\bf 4},
460-462 (1957).

Zeldovich, Ya. B.,  Ruzmaikin, A. A. \& Sokoloff, D. D., {\it
Magnetic  Fields in Astrophysics}, Gordon and Breach, New York
(1983).

Zeldovich, Ya. B.,  Ruzmaikin, A. A. \& Sokoloff, D. D.,  {\it The
Almighty Chance}, Word Scientific Publ., Singapore (1990).

\end{document}